\documentclass[11pt]{article}
\usepackage{graphicx} 

\usepackage[a4paper, margin=1in]{geometry}
\usepackage{setspace}

\usepackage[giveninits=true, maxcitenames=2]{biblatex} 
\renewbibmacro{in:}{}

\usepackage{rest-api}

\usepackage{amsthm}
\usepackage{amsmath}
\usepackage{amssymb}

\usepackage{xurl}
\usepackage{hyperref}

\usepackage{tikz}
\usetikzlibrary{arrows,fit,positioning,shapes,chains,fit,shapes,calc,backgrounds}
\usepackage{subcaption}
\usepackage{tabularx}

\theoremstyle{definition}

\usepackage{authblk}

\usepackage{orcidlink}

\title{\emph{MATWA}: A Web Toolkit for Matching under Preferences\thanks{Frederik Glitzner is supported by a Minerva Scholarship from the School of Computing Science, University of Glasgow. David Manlove is supported by the Engineering and Physical Sciences Research Council, grant number EP/X013618/1.}}

\author[ ]{Frederik Glitzner \orcidlink{0009-0002-2815-6368} and David Manlove \orcidlink{0000-0001-6754-7308}}
\affil[ ]{School of Computing Science, University of Glasgow, Glasgow G12 8QQ, UK}
\affil[ ]{\normalfont \texttt{f.glitzner.1@research.gla.ac.uk, david.manlove@glasgow.ac.uk}}
\date{}

\begin{document}

\maketitle

\begin{abstract}
Matching markets, where agents are assigned to one another based on preferences and capacity constraints, are pervasive in various domains. This paper introduces \emph{MATWA} ({\tt https://matwa.optimalmatching.com}), a web application offering a rich collection of algorithms for fundamental problem models involving matching under preferences. \textit{MATWA} provides results and visualisations of matching algorithm outputs based on different methods for providing problem instances.

In this paper, we describe the features of the system, illustrating its usage for different problem models, and outlining the algorithm implementations that are supported. We also give evidence of usability testing and illustrate how the system was used to obtain new empirical results for a specific matching problem.

\textit{MATWA} is intended to be a resource for the community of researchers in the area of matching under preferences, supporting experimentation as well as aiding the understanding of matching algorithms.
\end{abstract}

\section{Introduction}

\subsection{Matching Markets}

Matching markets involve assigning agents to one another, subject to various criteria. Here, the term \emph{agent} is used loosely to mean any participant in a matching process, and could include commodities and human subjects. In many cases, the agents form two disjoint sets, and we seek to assign the agents in one set to those in the other.  Examples include assigning junior doctors to hospitals, pupils to schools, kidney patients to donors, and so on.

We primarily focus on the case that a subset of the agents has \emph{ordinal preferences} (henceforth \emph{preferences}) over a subset of the others.  That is, there is a notion of first choice, second choice, third choice, etc. These rankings might not always be complete or strictly ordered. For example, in a course allocation setting, a student might rank their most desired courses in strict order, followed by some less favourable ones in a tie (indicating that they are equally good), and not include undesirable ones in the ranking at all. Typically, there are other constraints in addition to the preference lists. For example, a student might not take more than a certain number of courses at a time, and the course might have an upper capacity on the number of students that can be admitted to it.

Applications of matching problems involving preferences can be very large in practice; for example, the National Resident Matching Program (NRMP) in the US filled nearly 40,000 junior doctor positions through its preference-based matching system in 2024 \cite{nrmp}. Economists have identified several problems that arise in decentralised, free-for-all markets, in which the agents can negotiate with one another directly to arrange assignments \cite{RX94,NRS08}. \emph{Centralised matching schemes} can avoid some of the inherent problems in free-for-all markets. These work along the following lines: the input data involving the agents and their preferences over one another are collected by a given deadline by a trusted central authority.  This third party then computes an optimal matching concerning the supplied preference lists and capacities, and any other problem-specific constraints. By participating in the process, the agents agree that the outcome is binding (that said, the properties of the matching should ensure that participants do not have an incentive to deviate and form private arrangements outside of the matching). The precise definition of an \emph{optimal matching} has many variations depending on the context. It could involve, for example, maximising the number of course allocations, giving the maximum number of school leavers their first-choice university, or ensuring that no junior doctor and hospital have an incentive to reject their assignment\/s and become matched together instead.

Much of the previous work on matching algorithms has focused on the design and analysis of efficient algorithms for the matching problems that underpin these centralised matching schemes.

\subsection{Contribution and Significance}

In this paper, we introduce the \emph{Matching Algorithm Toolkit Web Application} (shortened \emph{MATWA}, reachable at {\tt https://matwa.optimalmatching.com}), a web application that makes available over 40 different algorithms for multiple fundamental matching problem classes for research, teaching, and demonstration purposes. Developed over 24 years by 20 researchers and project students at the University of Glasgow's School of Computing Science, it started as a project to collate implementations of state-of-the-art algorithms in the field of matching under preferences. Many of these algorithms do not have easily available implementations (for example in the case of the Tan-Hsueh algorithm \cite{tanhsueh} or the algorithm to generate the popular matching switching graph \cite{MI08}, we are unaware of publicly available implementations). \textit{MATWA} overcomes this barrier and offers the possibility to easily and quickly test and compare different algorithms. Overall, this is the most extensive collection of algorithm implementations for matching under preference problems available to date. It provides a simple user-friendly online interface that lets users find matchings for instances of the {\sc Stable Marriage}, {\sc House Allocation}, {\sc Hospitals / Residents}, {\sc Stable Roommates}, and {\sc Student-Project Allocation} problem classes using a range of algorithms for each, upload or randomly generate instances, and show or download the results with relevant statistics and visualisations. Lastly, \emph{MATWA}'s API is openly exposed and can be accessed independently, however, the application was developed mainly with web interface users in mind.

\subsection{Research and Educational Applications}
\begin{figure}[!htb]
    \centering
    \begin{subfigure}{.99\textwidth}
        \includegraphics[width=\hsize]{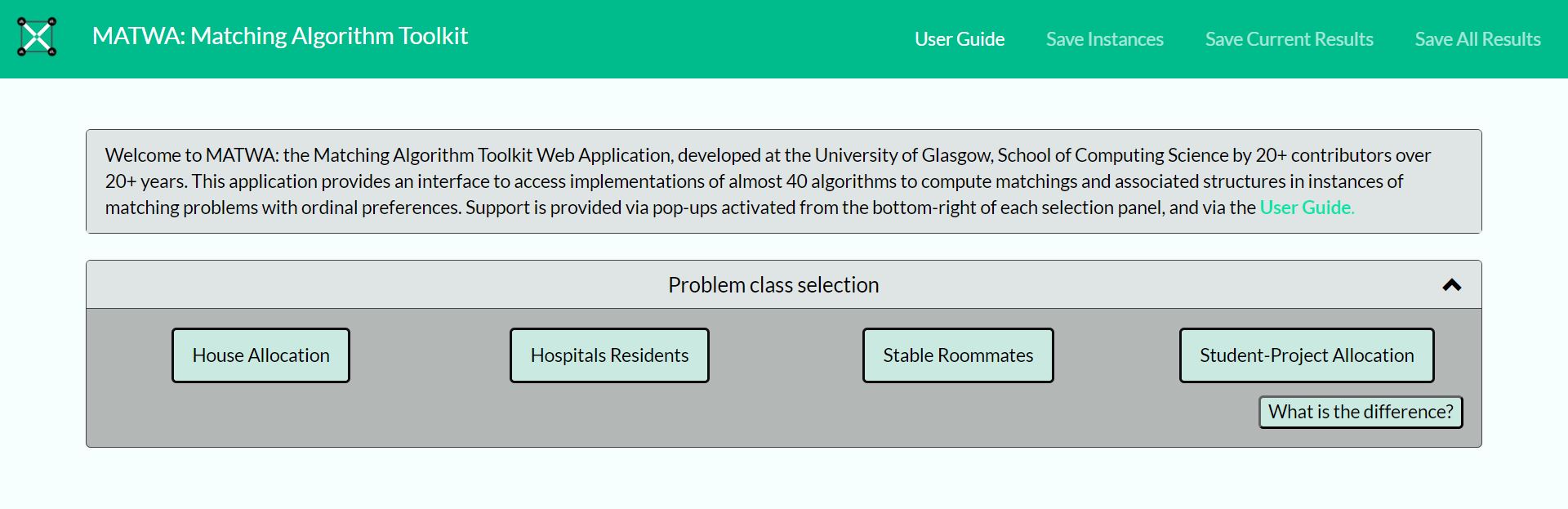}
    \end{subfigure}
    \caption{\textit{MATWA} Welcome Screen}
    \label{fig:welcome}
\end{figure}

There are many ways in which \textcite{matwa} can be used for research and for educational purposes. With a focus on usability during the development (see Figure \ref{fig:welcome} for the welcome screen), it is intended for researchers, students, and practitioners alike and can be used, for example, to easily and efficiently generate random problem instances with specific properties and obtain immediate feedback on different matching algorithms applied to them. For example, it can be used to demonstrate fairness in different matching mechanisms, to obtain an overview of the structural properties of a problem instance, through the computation of stable partitions in a {\sc Stable Roommates} instance, or by visualising the Hasse diagram of stable matchings and the rotation poset in a {\sc Stable Marriage} instance. Finally, summary statistics can be computed for different algorithms over many instances with regards to the size, cost, and profile of the respective matchings.

\subsection{Structure of the Paper}
The remainder of this paper is organised as follows. We will first give an overview of the different matching models and some natural objectives considered in the system in Section \ref{section:modelsobj}. Then, we will outline some typical use cases of \textcite{matwa}, describe the landscape of existing software for these, and introduce our system in Section \ref{section:software}. In Section \ref{section:features}, we will showcase the user experience and the available features of the front-end interface of our system. Following this, via a user study and an empirical evaluation conducted using \emph{MATWA}, we will present some evidence for the usability and research applicability of \emph{MATWA} in Section \ref{section:evaluation}. Finally, we will conclude and present some future work in Section \ref{section:conclusion}. Furthermore, Appendix \ref{section:apiappendix} contains an example usage of the public API endpoints and Appendix \ref{section:algorithms} lists all currently available algorithms.

\section{Models and Objectives}
\label{section:modelsobj}

\subsection{Matching Models}

In this section we describe the matching problem models that are supported by \emph{MATWA}.  An instance $I$ of a matching problem involves a set of agents (or items) $A$ (which might be partitioned into sets $A_1,\dots, A_k$ of different kinds such as students, projects, and lecturers), some collection $\succ$ of preference relations of some or all of the agents, and a capacity function $c : A'\longrightarrow \mathbb Z^+$ for some subset $A'$ of $A$.

We expect the preference relation to be derived from \emph{preference lists} for a subset $A''$ of $A$, in which each agent in $A''$ ranks an acceptable subset of other agents in some order. Instance $I$ is a \emph{no-tie} model if every preference relation is required to be strict. 

In general, we aim to find an \emph{assignment} $M$ which is a set of agent pairs $(a_i,a_j)$ such that $a_j$ belongs to $a_i$'s preference list and vice versa. If $A$ is partitioned into disjoint sets of agents of specific types, an assignment will normally comprise ordered pairs, otherwise it will normally be represented by unordered pairs of the form $\{a_i,a_j\}$. Furthermore, an assignment $M$ is a \emph{matching} if for all agents $a_k\in A'$, the set of agents $M(a_k)$ assigned to $a_k$ respects the capacity constraints, i.e. $\vert M(a_k)\vert \leq c(a_k)$. 

In a matching $M$, an agent $a_k\in A$ is said to be \emph{unassigned} if $M(a_k)=\varnothing$, otherwise it is said to be \emph{assigned}. Furthermore, given an agent $a_k\in A'$, if $\vert M(a_k)\vert < c(a_k)$, then $a_k$ is \emph{under-subscribed}, if $\vert M(a_k)\vert = c(a_k)$, then $a_k$ is \emph{full}, otherwise $a_k$ is \emph{over-subscribed}.

The system differentiates between different fundamental matching models, each of which covers similar configurations of the abstract model, all of which are outlined below. 

\paragraph{Hospitals / Residents ({\sc hr})}
The {\sc hr} problem \cite{GS62, GI89, RS90, Man08} (sometimes referred to as the College (or University or Stable) Admissions problem or the Stable Assignment problem) was first defined by Gale and Shapley in their seminal paper ``College Admissions and the Stability of Marriage'' \cite{GS62}. It arises, for example, when allocating junior doctors to hospitals.

An instance $I$ of {\sc hr} involves two disjoint sets of agents -- a set $R=\{r_1,\dots,r_{n_1}\}$ of \emph{residents} and a set $H=\{h_1,\dots,h_{n_2}\}$ of \emph{hospitals}. Each resident has capacity 1 and each hospital $h_j\in H$ has a positive integral capacity $c(h_j)$, indicating the number of \emph{posts} that $h_j$ offers. Each resident ranks an acceptable subset of hospitals in order of preference, and vice versa.

\paragraph{Stable Marriage ({\sc sm})}
The {\sc sm} problem \cite{GS62,Knu76,GI89,RS90,Irv08} was also first defined by \textcite{GS62} and is a restriction of the {\sc hr} problem where both sets of agents have equal cardinality and the capacity of every agent is 1. It arises, for example, when allocating mentees to mentors in a mentoring scheme. 

More formally, for an instance $I$ of {\sc sm}, $\vert R\vert =\vert H\vert$ and $c(h_j)=1$ for all $h_j\in H$.

\paragraph{House Allocation ({\sc ha})}
The {\sc ha} problem \cite{SS74,HZ79,DPS02,FSW03} (sometimes referred to as \emph{Capacitated House Allocation} {\sc (cha)} when the houses have capacity greater than 1) was defined by \textcite{HZ79} and is also a variant of the {\sc hr} problem where the set of hospitals does not have preferences over the residents. It arises, for example, when allocating a set of indivisible goods among a set of applicants.

An instance $I$ of {\sc ha} involves two disjoint sets of agents -- a set $A=\{a_1,\dots,a_{n_1}\}$ of \emph{applicants} and a set $H=\{h_1,\dots,h_{n_2}\}$ of \emph{houses}. Each applicant has capacity 1 and each house $h_j\in H$ either also has capacity 1 (in the case of {\sc ha}), or some positive integral capacity $c(h_j)$ (in the case of {\sc cha}). Each applicant ranks in order of preference a subset of the houses, whilst houses do not have preferences over applicants.

\paragraph{Stable Roommates ({\sc sr})}
The {\sc sr} problem \cite{GS62,Knu76,Irv85,GI89,RS90,IM02} (sometimes referred to as the Stable Matching problem \cite{AR94,Sub94,FMP00,Fle03}) was also first defined by \textcite{GS62} and is a generalisation of the {\sc sm} problem. It arises, for example, when allocating people to room-shares or players to doubles teams in sports tournaments.

An instance $I$ of {\sc sr} involves a single set of agents $A=\{a_1, \dots , a_n\}$. Each agent has capacity 1 and ranks in order of preference a subset of the other agents.

\paragraph{Student-Project Allocation ({\sc spa})}
The {\sc spa} problem \cite{ABRAHAM, SofiatPhD, CooperPhD, Augustine, sofiatStrong} can also be regarded as a generalisation of the {\sc hr} problem, with multiple versions studied in the literature. It arises, for example, when allocating students to final-year projects.

An instance $I$ of {\sc spa} generally involves three disjoint sets of agents -- a set $S=\{s_1,\dots,s_{n_1}\}$ of \emph{students}, a set $P=\{p_1,\dots,p_{n_2}\}$ of \emph{projects}, and a set $L=\{l_1,\dots,l_{n_3}\}$ of \emph{lecturers}. Each project is supervised by exactly one lecturer and every lecturer can offer multiple projects. Each student has capacity 1, whilst each project and lecturer has a positive integral capacity, indicating the number of students that can be assigned to it/them. Each student ranks in order of preference an acceptable subset of projects.

{\sc spa-s} is a variant of {\sc spa} in which lecturers also have preferences over an acceptable subset of students. Equivalently, we can consider the projects to have preferences over students inferred from their associated lecturer preferences.

\hspace{1em}

\begin{figure}[!htb]
    \centering

    \tikzstyle{block} = [rectangle, draw, fill=white, rounded corners,
                     minimum height=2em, minimum width =6em, text width=8em, align=center]
    \tikzstyle{line} = [draw, -latex']
    \begin{tikzpicture}[node distance = 2cm, auto]
    
    \node [circle, fill=black] (origin) {};
    \node [block, above right of=origin] (bip) {\footnotesize Bipartite};
    \node [block, below right of=origin, node distance=5cm] (nonbip) {\footnotesize Non-Bipartite};
    \node [block, above right of=bip, node distance=3cm] (onesided) {\footnotesize One-Sided Preferences};
    \node [block, below right of=bip, node distance=3cm] (twosided) {\footnotesize Two-Sided Preferences};
    \node [block, right of=onesided, node distance=5cm] (cha) {\footnotesize \sc Capacitated House Allocation};
    \node [block, above of=cha, node distance=1.5cm] (ha) {\footnotesize \sc House Allocation};
    \node [block, below of=cha, node distance=1.5cm] (spa) {\footnotesize \sc Student-Project Allocation};
    \node [block, below of=spa, node distance=1.5cm] (hr) {\footnotesize \sc Hospitals / Residents};
    \node [block, below of=hr, node distance=1.5cm] (sm) {\footnotesize \sc Stable Marriage};
    \node [block, below of=sm, node distance=1.5cm] (spas) {\footnotesize \sc Student-Project Allocation w/ Lecturer Preferences};
    \node [block, below of=spas, node distance=1.5cm] (sr) {\footnotesize \sc Stable Roommates};
    
    \path [line] (origin) -- (bip);
    \path [line] (origin) -- (nonbip);
    \path [line] (bip) -- (onesided);
    \path [line] (bip) -- (twosided);
    \path [line] (onesided) -- (ha);
    \path [line] (onesided) -- (cha);
    \path [line] (onesided) -- (spa);
    \path [line] (twosided) -- (hr);
    \path [line] (twosided) -- (sm);
    \path [line] (twosided) -- (spas);
    \path [line] (nonbip) -- (sr);
    \end{tikzpicture}
    \caption{High-Level Overview of Matching Problem Classes}
    \label{fig:problemoverview}
\end{figure}
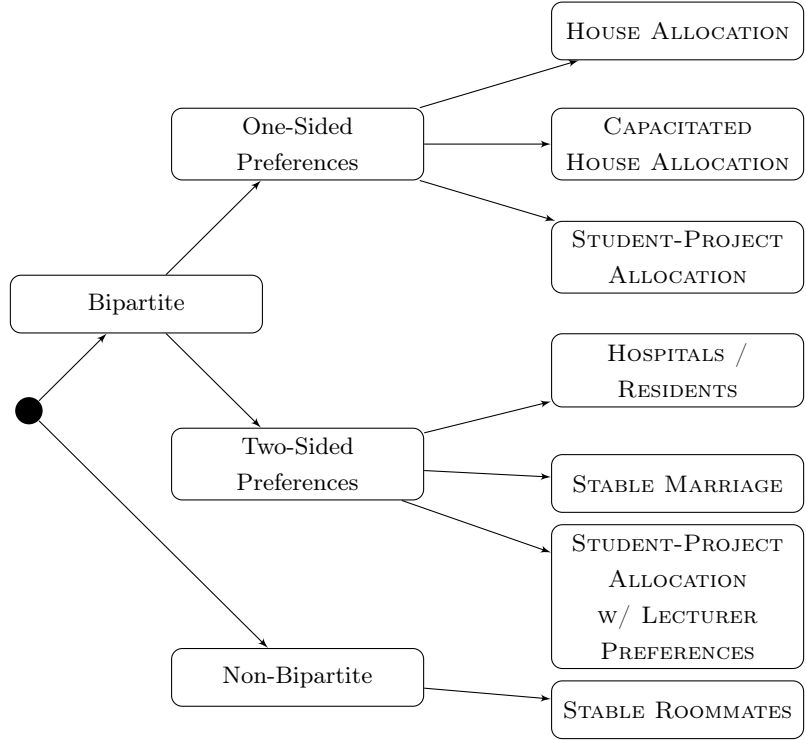

Figure \ref{fig:problemoverview} shows a high-level overview over the problem classes considered. We differentiate between bipartite and non-bipartite models, where in the former, we seek a matching between two disjoint sets of agents. In bipartite models, we also differentiate between one-sided and two-sided preferences, where in the former, only agents from one of the agent sets in the bipartition has preferences, whereas agents from both agent sets have preferences in the latter case.

\subsection{Objectives and Constraints}

\paragraph{Fundamental Solution Concepts}
In the literature, a range of different constraints and objectives have been studied for each problem class. Without going into the specifics of each of them, we will give an outline of some. 

We have already noted that a matching should satisfy all given capacity constraints. Generally, it is also often useful in practice to maximise the \emph{cardinality} (size) of the matching. Additional objectives stemming from cooperative game theory \cite{matchUp} are often considered. For example, a matching is \emph{Pareto-optimal} if there is no other matching in which at least one agent is better off and no agent is worse off. Similarly, a matching is \emph{popular} if there is no other matching that is preferred by a majority of the agents (who have a strict preference between the two matchings). Finally, a matching is \emph{stable} if there is no pair of agents who find each other acceptable such that each agent is either under-subscribed, or is full and prefers the other to their worst assigned partner.

\paragraph{Fairness and Optimality}

For a matching $M$ and for some subset $A_k\subseteq A''$ of agents that have preferences, many common optimality and fairness measures depend on the \emph{profile} $p^k(M)=(p_1^k \; p_2^k \dots p_{R}^k)$, a vector in which $p_i^k$ counts the number of agents from $A_k$ who are assigned in $M$ to someone of rank $i$ in their preference list, where $R$ is the maximum length of any agent's preference list. The \emph{regret} $r_k(M)$ of $M$ (relative to an agent set $A_k$)  is the largest $i$ such that $p_i^k$ is positive. Finally, the egalitarian \emph{cost} of $M$ is the weighted sum of its profile $c^k(M)=\sum_{1\leq i \leq R}i*p_i^k$. 

With the above measures, we can consider, for example, the problem of finding a stable matching with minimum egalitarian cost, or minimum regret. Alternatively, we could maximise the number of first choices satisfied in the assignment, or minimise the number of $R$th choices. Furthermore, we can aim for a hierarchical optimisation target such as a \emph{rank-maximal} or \emph{generous} profile. In the former, $p_1^k$ is maximal, subject to this, $p_2^k$ is maximal, and so on, up until $p_{R}^k$. In the latter, $p_{R}^k$ is minimal, subject to this, $p_{R-1}^k$ is minimal, and so on.

\section{Software Tools}
\label{section:software}

\subsection{Some typical use cases for \textit{MATWA}}

Some common computational problems associated with matching under preferences are to:
\begin{itemize}
    \item find a matching that is game-theoretically useful or economically fair (e.g., given an {\sc sm} instance, find a stable matching with minimum egalitarian cost);
    \item find a matching that approximately satisfies optimality criteria that are intractable to optimise for in general (e.g., given an instance of {\sc hr} with ties, find a (weakly) stable matching using Kir\'aly's approximation algorithm \cite{a6030471} (for the NP-hard problem of finding a maximum size (weakly) stable matching));
    \item study and compare the behaviour of different algorithms (e.g., given an {\sc ha} instance, compare the size, cost and profile of maximum Pareto optimal, popular and rank-maximal matchings);
    \item enumerate all matchings subject to some criteria (e.g., given an {\sc sm} instance,  find all stable matchings, view the rotation poset and the Hasse diagram for the set of stable matchings; for an {\sc sr} instance, find all stable matchings, view the rotation poset); and
    \item enumerate all pairs contained in some set of matchings (e.g., given an {\sc sm} instance, find all pairs contained in some stable matching, and similarly for {\sc sr}).
\end{itemize}

\textcite{matwa} addresses all of these problems and presents relevant statistics or visualisations for each result.

\subsection{Related Software}

Various software tools have been developed previously to solve instances of matching problems. Of course, the extensive literature published on algorithms for matching under preferences can be used for custom implementations, either from scratch or building on top of frameworks such as NetworkX \cite{networkx} or LEMON \cite{lemon}. In fact, these frameworks already offer algorithms to compute maximum cardinality matchings, and maximum- and minimum-weight matchings, but they do not provide implementations of preference-based algorithms. One could also use general-purpose integer programming or constraint programming solvers such as GLPK \cite{glpk} or CP-SAT \cite{cpsat} to manually implement problem models that offer similar outcomes to those provided by the specialised algorithms in the literature, although potentially much more inefficiently.

There are also various special-purpose implementations and source code repositories which offer specific algorithms for the {\sc cha, sm, hr, sr} and {\sc spa} problem classes. For example, the Python package {\tt Matching} \cite{matchingpackage} looks to be the most comprehensive with regards to stable matching implementations, offering one algorithm for each of {\sc sm}, {\sc hr}, {\sc sr}, and {\sc spa}. However, it is not possible to compare different algorithms for the same problem class and no alternative solution concepts are provided for instances that do not admit any stable matching. Code by \textcite{coopercode} (also available as Python package {\tt matchingproblems}) can also find matchings with specific optimality properties such as stability, maximum size, or profile-optimality, for an extended version of the {\sc spa-s} problem (which can also model other bipartite problems such as {\sc hr}). However, there are no visualisations and no graphical user interface.

As our goal is to provide a user interface and platform for quick verification of ideas and experimentation with problem instances, we will now focus only on other online applications for matching problems involving costs or preferences. An online tool by \textcite{VisualMatchingAlgo} lets users find a maximum cardinality matching, but does not offer algorithms satisfying preference-based objectives such as stability, profile-optimality, or popularity. Along similar lines, a web-based tool by \textcite{TUM} lets users find a minimum cost maximum cardinality matching in complete \emph{weighted bipartite graphs} (bipartite graphs where edges have real- or integer-valued costs) using the Hungarian method \cite{Hungarian}, also visualising the steps the algorithm takes, but without taking preference-based objectives into account.

Other educational tools such as {\tt MatchU}, due to \textcite{MatchU}, provide some fundamental matching algorithms for {\sc sm, hr}, and {\sc ha} instances in their tool and visualise executions of the algorithms step-by-step. Specifically, it makes available the classical Gale-Shapley algorithm \cite{GS62} for {\sc sm}, its extension to {\sc hr}, and multiple mechanisms for {\sc ha}. It is beginner-friendly and does not assume advanced knowledge of the topic, gives presets for instances, but also lets the user modify the instances manually. However, the tool is not primarily intended for research and, as such, only permits small instances due to display constraints, does not allow the user to upload specific instances or randomly generate instances with specific characteristics (just random generation with fixed presets of parameters is supported), download results, or compare the results of different algorithms applied to the same problem instance. Similar features are offered by the Matching Algorithm Visualiser ({\tt AlgMatch}), due to \textcite{AlgMatch}, which can also find stable matchings and visualise the algorithmic steps in a user-friendly way, but similar restrictions apply. 

Specific to {\sc cha}, a tool by \textcite{omalleytool} can compute various profile-optimal matchings on instances pasted into text boxes but does not consider problems with two-sided preferences or allow random instances. Similarly, \textcite{RichardMorey} provided a special-purpose web application that lets the user upload instances in separate files and uses an implementation of the student-oriented stable matching algorithm \cite{ABRAHAM} for {\sc spa-s} instances with two-sided preferences. However, it has no visualisations, or random instance generator, and is therefore not suitable for experimentation.

Table \ref{table:solvers_software} shows an overview of these existing web-based solvers in comparison to \textcite{matwa}. Columns 2-5 indicate whether the tool supports the respective problem class, Column 6 indicates the total number of algorithms offered by the tool, Column 7 indicates whether the tool can generate random instances, and Column 8 indicates whether the tool can be used to export its matching results. Clearly, \textit{MATWA} offers the highest problem class coverage, far exceeds the number of algorithms offered by any other application, and also supports the final two features. 

\begin{table}[!htb]
    \centering
    \small
    \begin{tabularx}{.95\textwidth}{c|c|c|c|c|c|c|c}
    \multicolumn{1}{p{2.5cm}|}{\centering \textbf{Tool/\\Author} } & 
    \multicolumn{1}{p{.8cm}|}{\centering \textbf{\sc cha}} & 
    \multicolumn{1}{p{.8cm}|}{\centering \textbf{\sc hr}} & 
    \multicolumn{1}{p{.8cm}|}{\centering \textbf{\sc sr}} & 
    \multicolumn{1}{p{.8cm}|}{\centering \textbf{\sc spa}} & 
    \multicolumn{1}{p{1.8cm}|}{\centering \textbf{\#Algos} } &  
    \multicolumn{1}{p{2.1cm}|}{\centering \textbf{Random\\Instances} } & 
    \multicolumn{1}{p{2.05cm}}{\centering \textbf{Export\\Feature} } \\\hline
    MATWA & \cellcolor[HTML]{b3ffb3}Y & \cellcolor[HTML]{b3ffb3}Y & \cellcolor[HTML]{b3ffb3}Y & \cellcolor[HTML]{b3ffb3}Y & \cellcolor[HTML]{b3ffb3}41 & \cellcolor[HTML]{b3ffb3}Y & \cellcolor[HTML]{b3ffb3}Y \\\hline
    MatchU \cite{MatchU} & \cellcolor[HTML]{b3ffb3}Y &\cellcolor[HTML]{b3ffb3}Y &\cellcolor[HTML]{ff9999}N &\cellcolor[HTML]{ff9999}N & 6 & \cellcolor[HTML]{ff9999}N & \cellcolor[HTML]{ff9999}N \\\hline 
    \textcite{omalleytool} & \cellcolor[HTML]{b3ffb3}Y & \cellcolor[HTML]{ff9999}N & \cellcolor[HTML]{ff9999}N & \cellcolor[HTML]{ff9999}N & 6 & \cellcolor[HTML]{ff9999}N & \cellcolor[HTML]{ff9999}N \\\hline
    AlgMatch \cite{AlgMatch} & \cellcolor[HTML]{ff9999}N & \cellcolor[HTML]{b3ffb3}Y & \cellcolor[HTML]{b3ffb3}Y & \cellcolor[HTML]{b3ffb3}Y &  6 & \cellcolor[HTML]{b3ffb3}Y & \cellcolor[HTML]{ff9999}N \\\hline
    \textcite{RichardMorey} & \cellcolor[HTML]{ff9999}N &\cellcolor[HTML]{ff9999}N &\cellcolor[HTML]{ff9999}N &\cellcolor[HTML]{b3ffb3}Y &  \cellcolor[HTML]{ff9999}1 & \cellcolor[HTML]{ff9999}N & \cellcolor[HTML]{b3ffb3}Y \\\hline
    TUM \cite{TUM} & \cellcolor[HTML]{ff9999}N & \cellcolor[HTML]{ff9999}N & \cellcolor[HTML]{ff9999}N & \cellcolor[HTML]{ff9999}N & 3 & \cellcolor[HTML]{b3ffb3}Y & \cellcolor[HTML]{b3ffb3}Y  \\\hline
    \textcite{VisualMatchingAlgo} & \cellcolor[HTML]{ff9999}N & \cellcolor[HTML]{ff9999}N & \cellcolor[HTML]{ff9999}N & \cellcolor[HTML]{ff9999}N & 5 & \cellcolor[HTML]{b3ffb3}Y & \cellcolor[HTML]{ff9999}N
    \end{tabularx}
    \caption{Comparison of Web-Based Software}
    \label{table:solvers_software}
\end{table}

\subsection{Our System and Contributors}
Our software, \textcite{matwa}, is a general-purpose web application developed over time within the University of Glasgow's School of Computing Science by at least 20 previous contributors over a period of 24 years. Most contributions were part of degree projects dedicated to developing parts of the system, many still in use, some discontinued, and some refactored. With almost 500 Java classes in the back-end code alone, this application has grown into a substantial software project accumulating many state-of-the-art matching algorithms that are not publicly available, to the best of our knowledge (such as the Tan-Hsueh algorithm \cite{tanhsueh} or the popular matching switching graph algorithm by \textcite{IM08}).

The project started as a command-line tool to make matching algorithms available in a more standardised way. \textcite{AlesRemta} designed and implemented a corresponding API framework. \textcite{PhilipYuile} created the first Graphical User Interface to expose many of the algorithms offered by the command-line toolkit. Later, \textcite{Boris} built a Django front-end web application and extended the existing back-end. Several other students and researchers have since updated and extended the application, for example by adding more algorithm implementations, the {\sc spa} problem class, and integrating more graphs and visualisations (see \textcite{matwamanual} for a full contributor list).

Currently, the system consists of independent front- and back-end services, which are both hosted on an internal University of Glasgow server. The front-end uses basic bootstrap components, and JavaScript and jQuery for the interactions. The back-end is more advanced, handles most of the application logic, and is implemented in Java as a Spring Boot-based REST API (for an example showing how to use the API without the front-end, see Appendix \ref{section:apiusage}).

\section{Features and User Experience}
\label{section:features}

This section focuses on the front-end interface of \textcite{matwa}, but the principles regarding input, algorithms, and output also apply to the API endpoints.

\subsection{UX Flow}

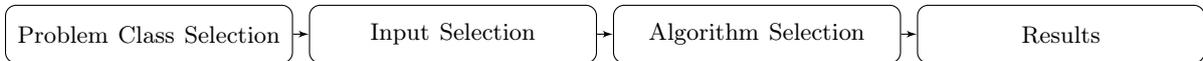
\begin{figure}[!htb]
    \centering

    \tikzstyle{block} = [rectangle, draw, fill=white, rounded corners,
                     minimum height=2em, minimum width =6em, text width=9.2em, align=center]
    \tikzstyle{line} = [draw, -latex']
    \begin{tikzpicture}[node distance = 2.4cm, auto]
    
    \node [block] (problem) {\footnotesize Problem Class Selection};
    \node [block, right of=problem, node distance=4.cm] (input) {\footnotesize Input  Selection};
    \node [block, right of=input, node distance=4.cm] (algo) {\footnotesize Algorithm Selection};
    \node [block, right of=algo, node distance=4.cm] (results) {\footnotesize Results};
    
    
    \path [line] (problem) -- (input);
    \path [line] (input) -- (algo);
    \path [line] (algo) -- (results);
    
    \end{tikzpicture}
    
    \caption{Basic User Experience Flow}
    \label{fig:ux_flow}
\end{figure}

The basic user experience flow can be seen in Figure \ref{fig:ux_flow}. From the landing page, the user selects one of the fundamental problem classes and then chooses the method for communicating data about the instance to the system. Subsequently, the provided or generated instance is then processed and a list of available algorithms is presented. After the user chooses the desired algorithms, the system computes the results and presents them separately by algorithm. Note that the user can also go back to any previous step at any point. 

\subsection{Input}
There are three different ways to provide problem instance data to the application as input. There is a text box to type or paste custom instances, a mechanism to upload instances via a text file, and a random generator form in which the user can specify parameter values (such as the numbers of agents of different types, the lengths of preference lists and the total capacities) and the system randomly generates such an instance. For the first two input types, a strict format must be followed which can be found in the \textcite{matwamanual}. As an example, Figure \ref{fig:spaformat} shows the schema for a {\sc spa-s} instance, Figure \ref{fig:formatexample} shows such an instance with 3 students, 4 projects, and 2 lecturers, and Figure \ref{fig:ties_example} shows a {\sc spa} instance without lecturer preferences over students, and some ties in the student preference lists indicated by the parentheses. Specifically, the first line indicates the number of students, projects, and lecturers in each example, followed by the student preference lists over projects (e.g. student 2 only finds projects 2 and 3 acceptable and prefers the former over the latter). The next block of sequentially numbered lines indicate the lecturer capacities and preferences over students (if any), and, finally, the last set of lines gives the project ids, their capacities, and the corresponding supervising lecturers.

\begin{figure}[!htb]
  
    \pgfdeclarelayer{background}
    \pgfsetlayers{background,main}
    
    \tikzstyle{vertex}=[circle,fill=black!25,minimum size=20pt,inner sep=0pt]
    \tikzstyle{edge} = [draw,thick,-]
    \tikzstyle{weight} = [font=\small]
    \tikzstyle{selected edge} = [draw,line width=5pt,-,red!50]
    
    \centering
    \footnotesize
    
    \begin{minipage}{0.33\linewidth}
        \#Students \#Projects \#Lecturers\\
        StuId: Pref1 Pref2 (...)\\
        StuId: Pref1 Pref2 (...)\\
        StuId: Pref1 Pref2 (...)\\
        LecId: LecCapacity: Pref1 Pref2 (...)\\
        LecId: LecCapacity: Pref1 Pref2 (...)\\
        ProjId: ProjCapacity: LecId\\
        ProjId: ProjCapacity: LecId\\
        ProjId: ProjCapacity: LecId\\
        ProjId: ProjCapacity: LecId
        \caption{{\sc spa-s} Input Schema}
        \label{fig:spaformat}
     \end{minipage}\hfill\begin{minipage}{0.33\linewidth}\hspace*{15mm}
        3 4 2\\\hspace*{15mm}
        1: 1 2\\\hspace*{15mm}
        2: 2 3\\\hspace*{15mm}
        3: 1 3\\\hspace*{15mm}
        1: 2: 1 2 3\\\hspace*{15mm}
        2: 1: 2 1 3\\\hspace*{15mm}
        1: 1: 1\\\hspace*{15mm}
        2: 2: 1\\\hspace*{15mm}
        3: 2: 2\\\hspace*{15mm}
        4: 1: 2
        \caption{{\sc spa-s} Instance}
        \label{fig:formatexample}
    \end{minipage}\hfill\begin{minipage}{0.33\linewidth}\hspace*{15mm}
        3 4 2\\\hspace*{15mm}
        1: (1 2)\\\hspace*{15mm}
        2: 2 3\\\hspace*{15mm}
        3: (1 3)\\\hspace*{15mm}
        1: 2:\\\hspace*{15mm}
        2: 1:\\\hspace*{15mm}
        1: 1: 1\\\hspace*{15mm}
        2: 2: 1\\\hspace*{15mm}
        3: 2: 2\\\hspace*{15mm}
        4: 1: 2
        \caption{{\sc spa} Instance}
        \label{fig:ties_example}
    \end{minipage}
\end{figure}

With regards to the randomly generated instances, depending on the problem class, there are up to 14 parameters that can be specified, with default values provided on opening. Some common themes in the parameter selection across the problem classes are:
\begin{itemize}
    \item \textbf{Number of agents:} always provided per agent set;
    \item \textbf{Capacity of agents:} only for agent sets that allow a capacity higher than 1;
    \item \textbf{Probability of ties:} probability for a preference list entry to be tied with its successor;
    \item \textbf{Skewness:} difference in popularity between the most popular and the least popular agent in the set, with a linear distribution in between them;
    \item \textbf{Preference list lengths:} creates subsets of acceptable agents in the preference relation, with cardinality lying between specified lower and upper bounds between which the generator varies randomly; and
    \item \textbf{Even position distribution:} divides the total capacities equally within the corresponding set if checked, otherwise randomly splits them.
\end{itemize}

Note that {\sc sm} instances are automatically recognised as a specific kind of {\sc hr} instance in the system. Furthermore, the system also works with multiple instances provided or generated at once to get summary statistics, but, currently, fewer algorithms are available in this case for technical reasons. For a full list of parameters by problem class, see the \textcite{matwamanual}.

\subsection{Algorithm Selection}

\begin{figure}[!htb]
    \centering
    \begin{subfigure}{.99\textwidth}
        \includegraphics[width=\hsize]{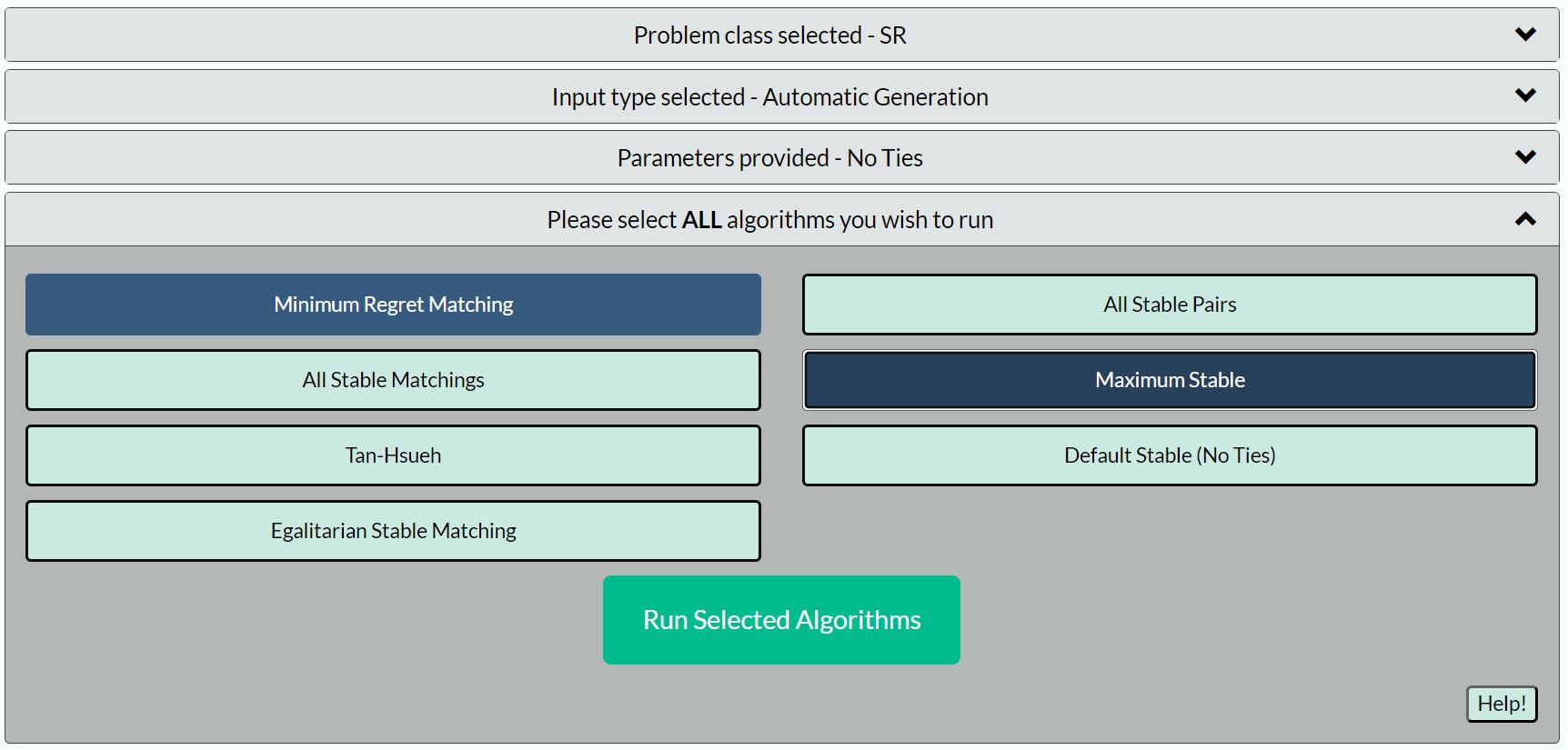}
    \end{subfigure}
    \caption{{\sc sr} Algorithm Selection}
    \label{fig:algoselection}
\end{figure}

After submitting the input, the system processes and analyses the problem instance and responds with a list of all applicable algorithms. The user can then choose between these and select all desired ones (see Appendix \ref{section:algorithms} for a full list of currently available algorithms). It is possible to select different types of algorithms at once, such as enumeration and visualisation algorithms, and other exact and approximation algorithms. Figure \ref{fig:algoselection} shows the algorithm selection screen for an {\sc sr} instance with complete preference lists and no ties.

\subsection{Output}
\begin{figure}[!htb]
    \centering
    \begin{subfigure}{.99\textwidth}
        \includegraphics[width=\hsize]{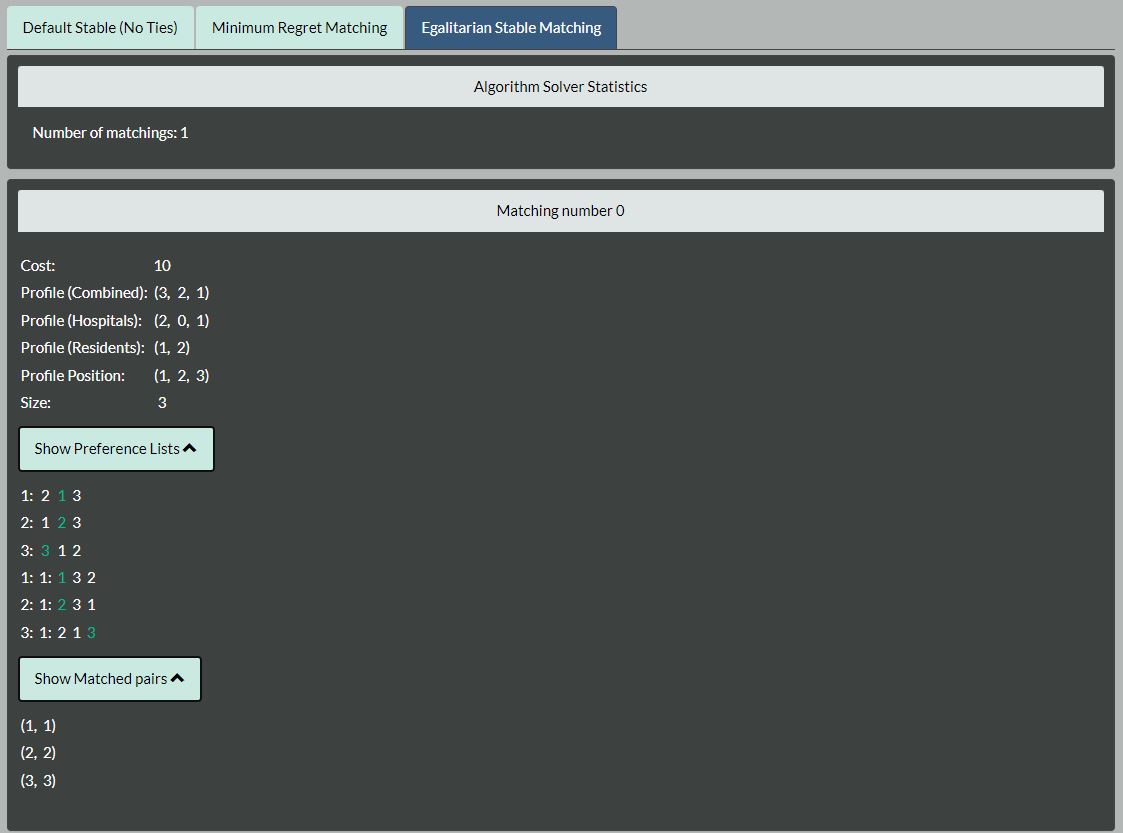}
    \end{subfigure}
    \caption{{\sc sm} Example Matching Output}
    \label{fig:matchingoutput}
\end{figure}

The results are provided in individual tabs for each algorithm. Generally, in each tab, the number of matchings (or other structure types sought, such as stable partitions) found is shown, which is 1 for all non-enumeration algorithms. Then, each matching is shown both as a set of pairs and as a highlighted set of entries in the preference lists. The statistics and further visualisations shown depend on the problem class and algorithm, but can include the cardinality, egalitarian cost, and profile, each of the latter two divided by agent group or in total. Figure \ref{fig:matchingoutput} shows an example output for an egalitarian stable matching of an {\sc sm} instance with 6 agents. 

\begin{figure}[!htb]
    \centering
    \begin{subfigure}{.8\textwidth}
        \includegraphics[width=\hsize]{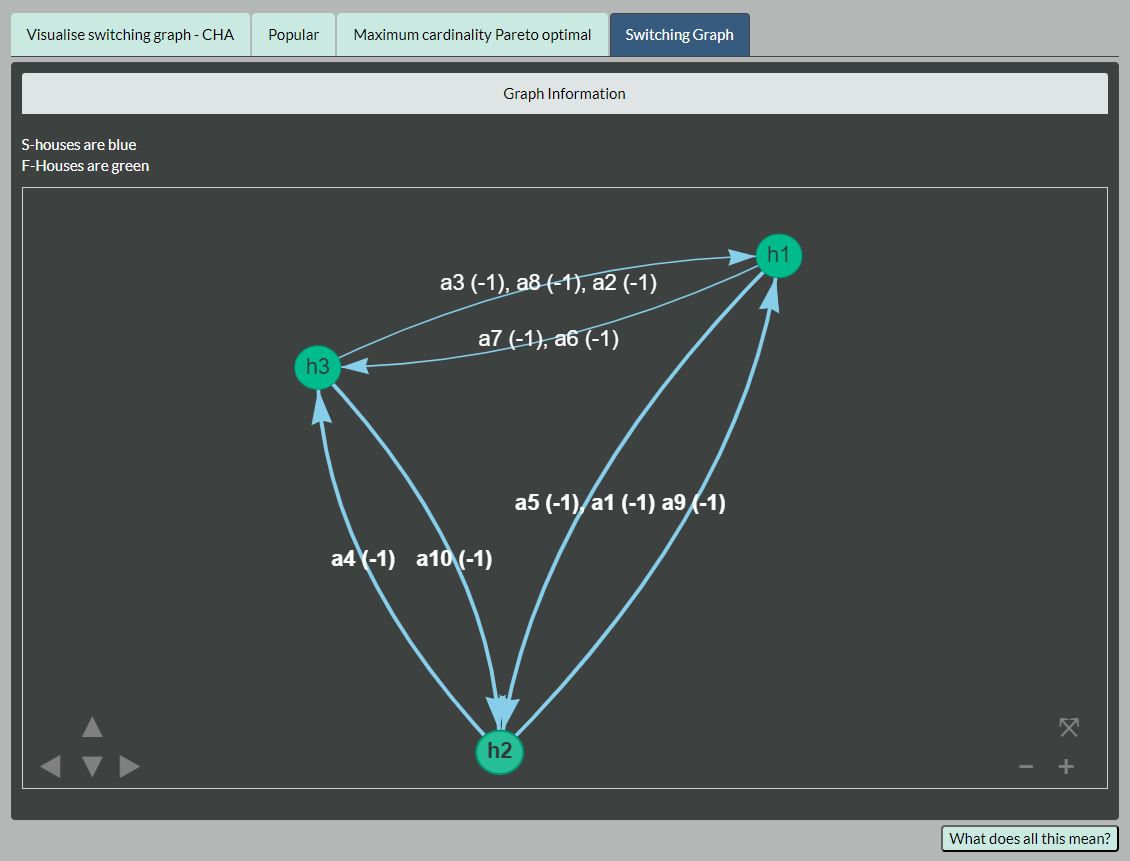}
    \end{subfigure}
    \caption{{\sc cha} Switching Graph}
    \label{fig:havis}
\end{figure}

\begin{figure}[!htb]
    \centering
    \begin{subfigure}{.32\textwidth}
        \includegraphics[width=\hsize]{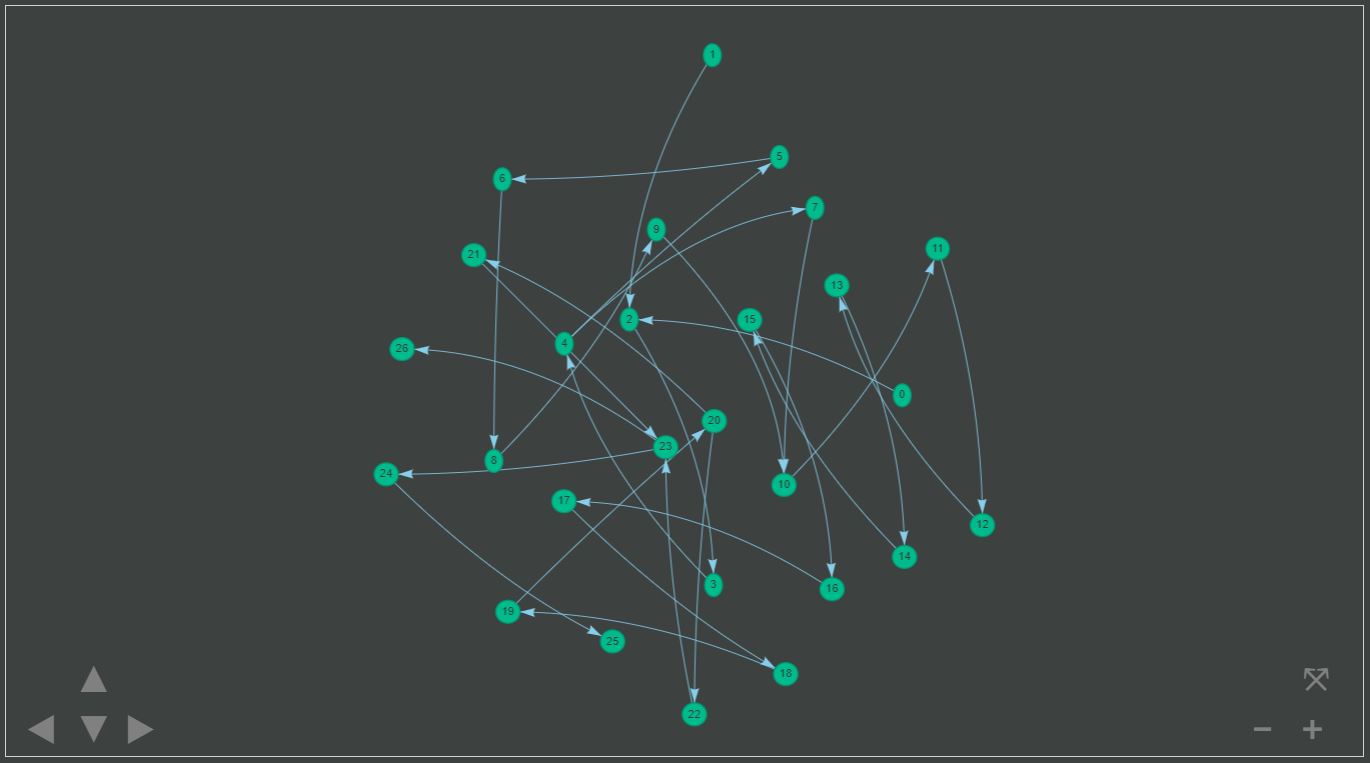}
    \end{subfigure}
    \begin{subfigure}{.32\textwidth}
        \includegraphics[width=\hsize]{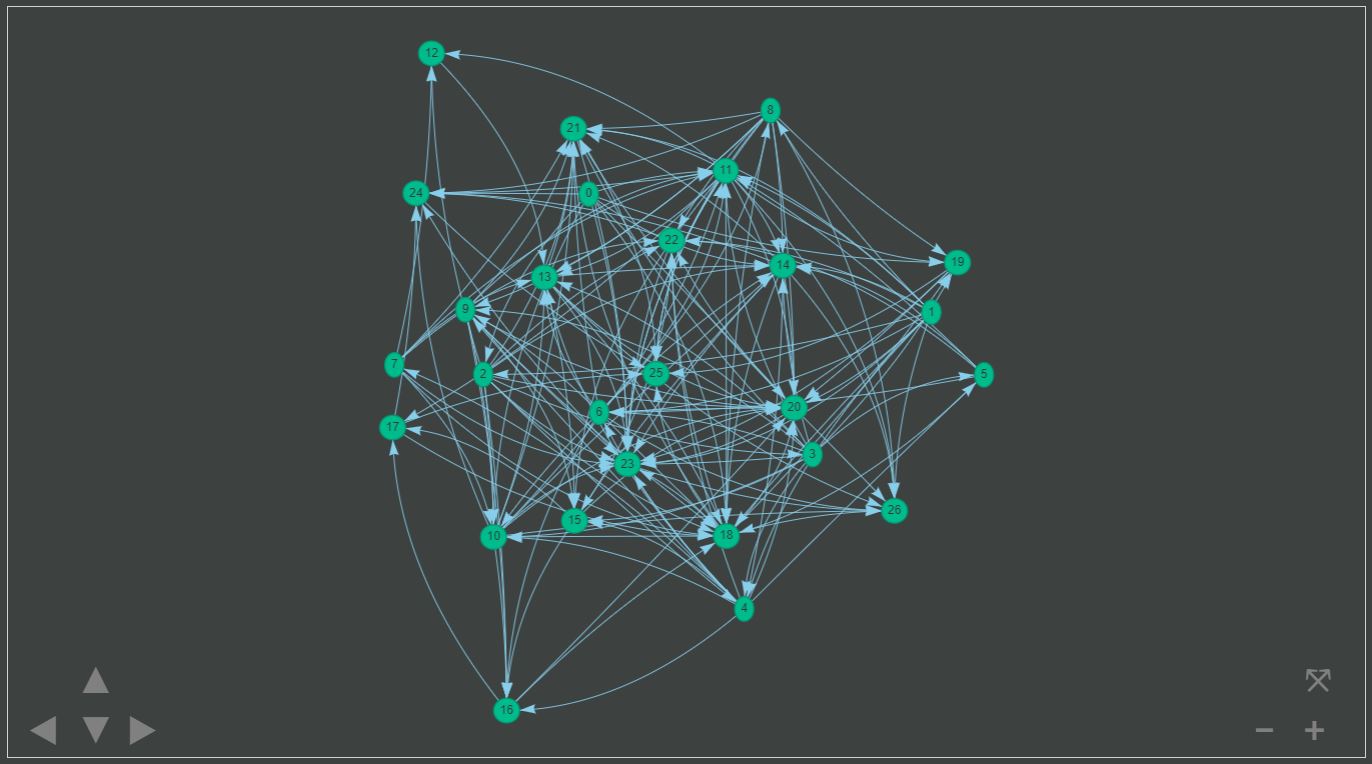}
    \end{subfigure}
    \begin{subfigure}{.32\textwidth}
        \includegraphics[width=\hsize]{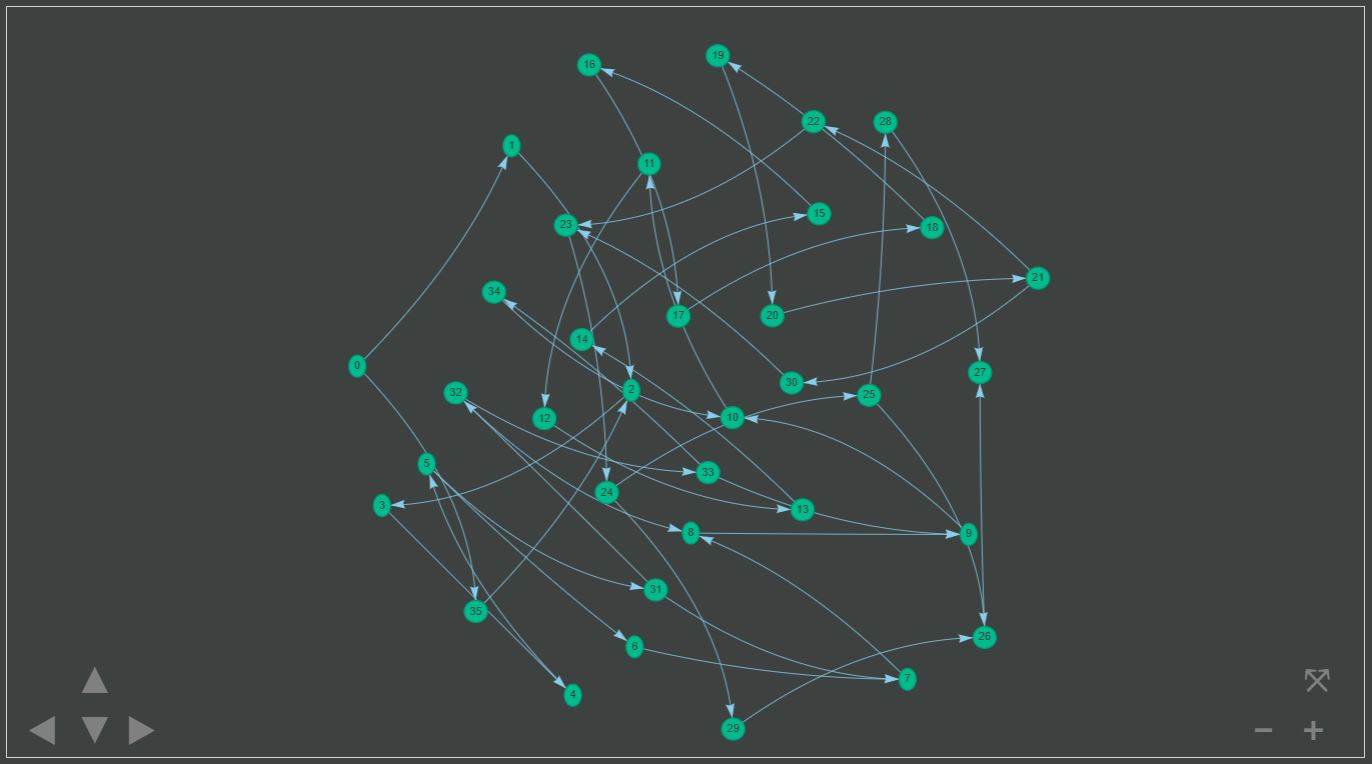}
    \end{subfigure}
    \caption{{\sc sm} Visualisations}
    \label{fig:smvis}
\end{figure}

Some of the visualisations available include \emph{switching graphs} for instances of the {\sc cha} problem class, \emph{rotation posets} for instances of the {\sc sr} and {\sc sm} problem classes, and \emph{rotation digraphs} and \emph{Hasse diagrams} for instances of the {\sc sm} problem class. The switching graph for a {\sc cha} instance with house capacities larger than one is a compact representation of its popular matchings \cite{McDermid2011}, with an example shown in Figure \ref{fig:havis}. The rotation poset, rotation digraph, and Hasse diagrams are compact representations of the instance's stable matchings \cite{matchUp}, with one example each (in order) shown in Figure \ref{fig:smvis}. 

The system's output can also be exported and saved. For example, randomly generated instances can easily be downloaded in text files, which can in turn be uploaded again to replicate the experiments. Furthermore, it is possible to download the generated matchings and their statistics, either separately or collated, as interactive stand-alone html files. Visualisations can also be downloaded, but they will not remain interactive.

\section{Evaluation}
\label{section:evaluation}

\subsection{Functional Testing}

There are automated functional and unit testing suites as well as manual pre-release integration testing procedures in place for \textcite{matwa}. With regards to automated tests, the test suites mainly cover API call and response handling, stability testing for stable matching algorithms, and other content verifications for API responses at various points of the UX flow.

\subsection{User Evaluation}
To evaluate the system, obtain feedback on its general usability, and gather suggestions for future improvements from potential users of the web application, a user study was carried out on an earlier version of the system by \textcite{frederikbsc}. 

The participants were informed of some basic features of the system and the {\sc spa} problem class specifically, as this was added by \textcite{frederikbsc}. Then, they were given structured tasks to complete such as generating and uploading instances, running different algorithms, and comparing and saving the results. Afterwards, the participants were asked to rate multiple usability-related questions on a Likert scale and to give open-ended answers on what they liked most about their interaction with the system, and what could be improved to enhance the usability or feature set of the system. Some usability-related questions are phrased the same as in a user study conducted on an even earlier version of the system by \textcite{Boris}, to enable the results of both user studies to be compared.

The 15 participants had varying levels of knowledge with regard to matching algorithms (mostly none to limited) and different academic positions (mostly undergraduate and PGR). Table \ref{fig:likert} presents the average responses to Likert-scale questions, rated between strongly disagree (one) and strongly agree (five).

\begin{table}[!htb]
    \centering
    \footnotesize
    \begin{tabularx}{.9\textwidth}{Xll}
        \textbf{Question}                                                     & \textbf{Median} & \textbf{Mean} \\
        The navigation to the previous selection tabs was intuitive.          & Strongly Agree  & 4.40          \\
        The input fields were working well and the input types were appropriate.   & Strongly Agree  & 4.67          \\
        The instance saving worked smoothly.                                  & Strongly Agree  & 4.93          \\
        The instance uploading worked smoothly.                               & Strongly Agree  & 5.00          \\
        The saving process worked smoothly.                                   & Strongly Agree  & 4.86          \\
        The results in the downloaded file show what I expected them to show. & Strongly Agree  & 4.71          \\
        The application was easy to navigate.                                 & Strongly Agree  & 4.93          \\
        Extracting information from the results was easy.                     & Strongly Agree  & 4.53          \\
        Overall, the tasks were easy to complete.                             & Strongly Agree  & 4.60         
    \end{tabularx}
    \caption{User Responses to Likert-scale Questions}
    \label{fig:likert}
\end{table}

The feedback roughly matches the results previously obtained by \textcite{Boris}, with a slight increase in scores for usability, which was the main emphasis of the work conducted in between the two versions. Multiple other helpful comments provided during the study were taken into account when developing the current version of the system.

\subsection{Example of Research Use}
\label{section:researchuse}

As part of an undergraduate research project, \textcite{frederikbsc} used the \textcite{matwa} system to evaluate five different algorithms for the {\sc spa} and {\sc spa-s} problem classes on more than 10,000 randomly generated problem instances in total. For all experiments, the instance sizes were parametrized by a factor $x\in\mathbb{N}$ such that in each instance, there were $5x$ students, $7x$ projects, and $x$ lecturers, and the total project and lecturer capacities were $10x$ and $7x$, respectively. Also, all preference lists were generated uniformly at random and none of the instances admitted ties.

\begin{figure}[!htb]
    \centering
    \begin{subfigure}{.495\textwidth}
        \includegraphics[width=\hsize]{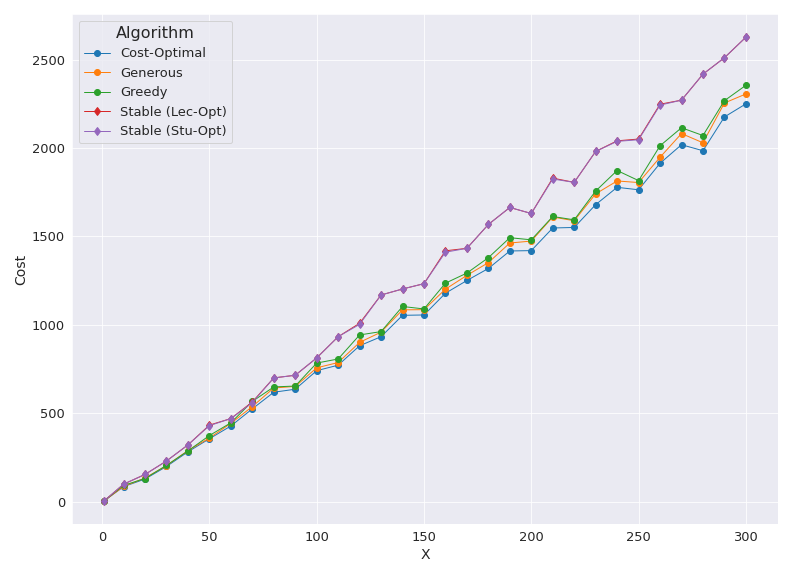}
        \caption{Total Student Cost}
        \label{fig:totalstucost_size}
    \end{subfigure}
    \begin{subfigure}{.495\textwidth}
        \includegraphics[width=\hsize]{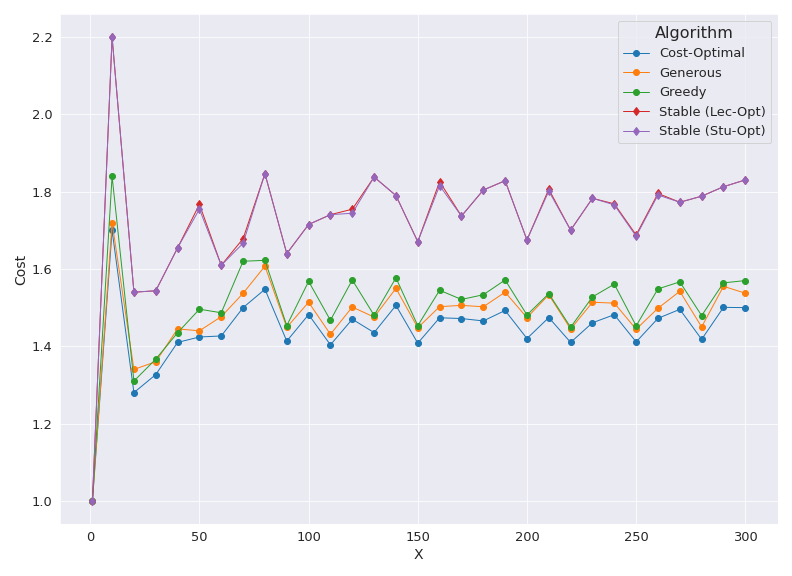}
        \caption{Average Student Cost}
        \label{fig:avgstucost_size}
    \end{subfigure}
    \caption{Student Costs as the Instance Size Varies}
    \label{fig:stucost_size}
\end{figure}

The first experiment investigated how the matching sizes, egalitarian costs per agent group, and computation times varied across the matchings computed by the different algorithms as the size of randomly generated instances grew to demonstrate the feasibility and utility of applying these algorithms in practice. Both the student- and lecturer-optimal stable matching algorithms \cite{ABRAHAM} as well as three profile-optimal algorithms \cite{Augustine} that only consider student preferences were evaluated to compare key measures when prioritising only student preferences over the aim to find a stable solution. As can be seen in Figure \ref{fig:stucost_size}, there is a strict hierarchy of algorithms maintained across the instance sizes with regard to the total and average student costs, with the highest cost achieved by the stable matching algorithms as expected. However, the average student cost is only slightly higher than the student-profile optimal matchings computed, with nearly constant average student costs of 1.8 and 1.4, respectively. Furthermore, most instances considered admit only one or very few stable matchings, hence the average student cost of the student- and lecturer-optimal algorithms is visually nearly indistinguishable.

\begin{figure}[!htb]
    \centering
    \begin{subfigure}{.49\textwidth}
        \includegraphics[width=\hsize]{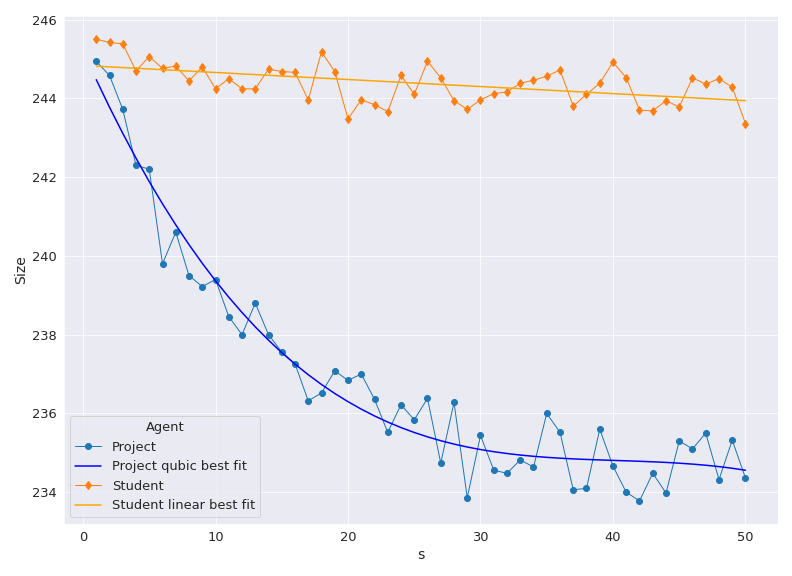}
        \caption{50 Lecturers}
        \label{fig:size_pop_50}
    \end{subfigure}
    \begin{subfigure}{.49\textwidth}
        \includegraphics[width=\hsize]{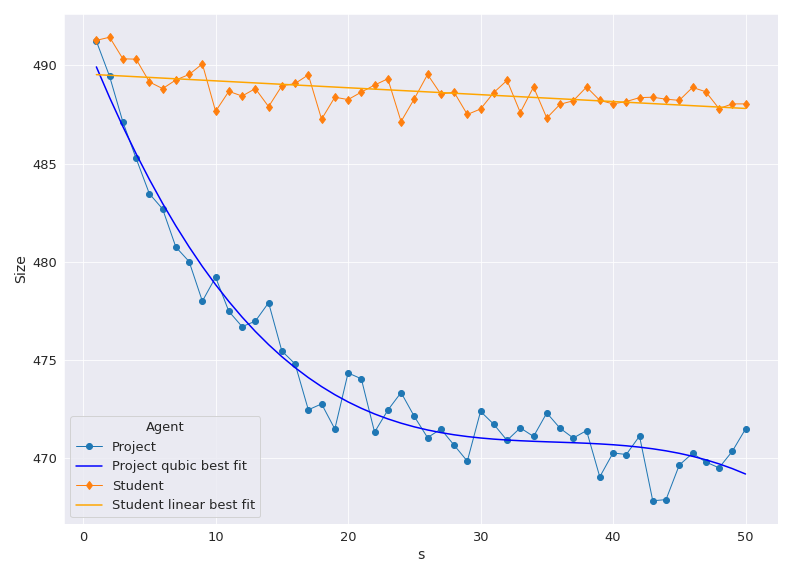}
        \caption{100 Lecturers}
        \label{fig:size_pop_100}
    \end{subfigure}
    \caption{Stable Matching Sizes as the Popularity Varies}
    \label{fig:size_pop}
\end{figure}

A second experiment investigated how the matching sizes varied across the matchings computed by the different algorithms as the agent popularity increased. Figure \ref{fig:size_pop} shows the average size (with lines of best fit) over instances with $x=50$ and $x=100$, where, when generating the preference lists from first choice to last choice for each agent, for each value of the skewness parameter $s$, the most popular project (or student) is $s$-times as likely to be chosen compared to the least popular project (or student). It was concluded that as the project popularity factor increased, the results suggested that the matching size decreased, but started to stabilise at a factor of $s=35$ for the instances considered. Furthermore, as the student popularity factor increased, there appeared to be a slight linear-like decrease in the matching size, although not as strong as when varying the project skewness. 

\section{Conclusion and Future Work}
\label{section:conclusion}

In conclusion, \textcite{matwa} is a comprehensive web application offering over 40 algorithms for various fundamental matching under preference problem models. It provides a valuable resource for demonstration, teaching, and research purposes through its user-friendly front-end interface and openly exposed back-end API. \textit{MATWA} is especially useful in facilitating easy testing, comparison, and analysis of different matching algorithms, which has been demonstrated by past research projects and usability studies performed on the system.

In the future, we plan to integrate more algorithms, for example, to generate all stable matchings for {\sc spa-s} instances, algorithms for the {\sc Stable Fixtures} problem class (a many-to-many extension of {\sc sr} \cite{IS07}), and to enumerate all stable partitions for {\sc sr} instances. Furthermore, we would like to extend the multi-instance support, implement more visualisations such as the meta-rotation posets \cite{metahr} for instances of the {\sc hr} problem class, provide different statistical distributions for the random instance generator, and output accurate runtimes for the execution of each algorithm.

\paragraph*{Acknowledgements} We would like to thank the anonymous MATCH-UP reviewers for their helpful suggestions. The full list of \textit{MATWA} contributors can be found at \url{https://matwa.optimalmatching.com/matching/manual}.


\printbibliography

@Article{McDermid2011,
author={McDermid, Eric
and Irving, Robert},
title={Popular matchings: structure and algorithms},
journal={Journal of Combinatorial Optimization},
year={2011},
month={10},
day={01},
volume={22},
number={3},
pages={339-358},
issn={1573-2886},
doi={10.1007/s10878-009-9287-9}
}

@misc{nrmp,
  title = {{2024 Main Residency Match by the Numbers}},
  author = {NRMP},
    note = {\url{https://www.nrmp.org/match-data/2024/06/results-and-data-2024-main-residency-match/} (accessed 2024-08-02)},
    year={2024}
}

@inproceedings{ACMM04,
  year = {2004},
  series = {Lecture Notes in Computer Science},
  title = {Pareto optimality in house allocation problems},
  pages = {3-15},
  author = {D.J. Abraham and K. Cechl\'arov\'a and D.F. Manlove and K. Mehlhorn},
  volume = {3341},
  publisher = {Springer},
  booktitle = {Proceedings of ISAAC '04: the 15th Annual International Symposium on Algorithms and Computation},
    doi={10.1007/978-3-540-30551-4_3}
}

@inproceedings{AIKM05,
  year = {2005},
  title = {Popular matchings},
  pages = {424-432},
  author = {D.J. Abraham and R.W. Irving and T. Kavitha and K. Mehlhorn},
  publisher = {ACM-SIAM},
  booktitle = {Proceedings of SODA '05: the 16th ACM-SIAM Symposium on Discrete Algorithms}
}

@article{AR94,
  year = {1994},
  title = {Stable matchings and linear inequalities},
  pages = {1-27},
  author = {H.G. Abeledo and U.G. Rothblum},
  journal = {Discrete Applied Mathematics},
  volume = {54},
    doi={10.1016/0166-218X(94)90130-9}
}

@article{DPS02,
  year = {2003},
  title = {On the complexity of equilibria},
  pages = {311-324},
  number = {2},
  author = {X. Deng and C. Papadimitriou and S. Safra},
  journal = {Journal of Computer and System Sciences},
  volume = {67},
doi = {10.1145/509907.509920},
}

@article{Fle03,
  year = {2003},
  title = {A fixed-point approach to stable matchings and some applications},
  pages = {103-126},
  number = {1},
  author = {T. Fleiner},
  journal = {Mathematics of Operations Research},
  volume = {28},
    doi={10.1287/moor.28.1.103.14256}
}

@article{FMP00,
  year = {2000},
  title = {A sublinear parallel algorithm for stable matching},
  pages = {297--308},
  number = {1--2},
  author = {T. Feder and N. Megiddo and S.A. Plotkin},
  journal = {Theoretical Computer Science},
  volume = {233},
doi = {10.1016/S0304-3975(99)00125-5}
}

@article{FSW03,
  year = {2003},
  title = {The complexity of economic equilibria for house allocation markets},
  pages = {219-223},
  author = {S.P. Fekete and M. Skutella and G.J. Woeginger},
  journal = {Information Processing Letters},
  volume = {88},
    doi={10.1016/j.ipl.2003.08.008}
}

@book{GI89,
  year = {1989},
  title = {The Stable Marriage Problem: Structure and Algorithms},
  author = {D. Gusfield and R.W. Irving},
  publisher = {MIT Press}
}

@InProceedings{hrt,
author="Irving, Robert
and Manlove, David
and Scott, Sandy",
title="The Hospitals/Residents Problem with Ties",
booktitle="Algorithm Theory - SWAT 2000",
year="2000",
publisher="Springer Berlin Heidelberg",
address="Berlin, Heidelberg",
pages="259--271",
doi={10.1007/3-540-44985-X_24},
isbn="978-3-540-44985-0"
}

@article{GS62,
  year = {1962},
  title = {College admissions and the stability of marriage},
  pages = {9-15},
  author = {D. Gale and L.S. Shapley},
  journal = {American Mathematical Monthly},
  volume = {69},
   doi = {10.2307/2312726}
}

@article{Gus87,
  year = {1987},
  title = {Three fast algorithms for four problems in stable marriage},
  pages = {111-128},
  number = {1},
  author = {D. Gusfield},
  journal = {SIAM Journal on Computing},
  volume = {16},
    doi={10.1137/0216010}
}

@article{gusfieldstructure,
author = {Gusfield, Dan},
title = {The Structure of the Stable Roommate Problem: Efficient Representation and Enumeration of All Stable Assignments},
journal = {SIAM Journal on Computing},
volume = {17},
number = {4},
pages = {742-769},
year = {1988},
doi = {10.1137/0217048}
}

@article{HZ79,
  year = {1979},
  title = {The efficient allocation of individuals to positions},
  pages = {293-314},
  number = {2},
  author = {A. Hylland and R. Zeckhauser},
  journal = {Journal of Political Economy},
  volume = {87},
    doi={10.1086/260757}
}

@article{IKMMP06,
  year = {2006},
  title = {Rank-maximal matchings},
  pages = {602-610},
  number = {4},
  author = {R.W. Irving and T. Kavitha and K. Mehlhorn and D. Michail and K. Paluch},
  journal = {ACM Transactions on Algorithms},
doi = {10.1145/1198513.1198520},
  volume = {2}
}

@article{ILG87,
  year = {1987},
  title = {An efficient algorithm for the ``optimal'' stable marriage},
  pages = {532-543},
  number = {3},
  author = {R.W. Irving and P. Leather and D. Gusfield},
  journal = {Journal of the ACM},
  volume = {34},
    doi={10.1145/28869.28871}
}

@article{IM02,
  year = {2002},
  title = {The {S}table {R}oommates {P}roblem with {T}ies},
  pages = {85-105},
  author = {R.W. Irving and D.F. Manlove},
  journal = {Journal of Algorithms},
  volume = {43},
    doi={10.1006/jagm.2002.1219}
}

@article{IM08,
  year = {2008},
  title = {Approximation algorithms for hard variants of the stable marriage and hospitals/residents problems},
  pages = {279-292},
  number = {3},
  author = {R.W. Irving and D.F. Manlove},
  journal = {Journal of Combinatorial Optimization},
  volume = {16}
}

@techreport{IMS02,
  year = {2002},
  title = {Strong stability in the {H}ospitals/{R}esidents problem},
  institution = {University of Glasgow, Department of Computing Science},
  number = {TR-2002-123},
  author = {R.W. Irving and D.F. Manlove and S. Scott},
  note = {Revised May 2005}
}

@incollection{Irv08,
  booktitle = {Encyclopedia of Algorithms},
  year = {2008},
  author = {R.W. Irving},
  publisher = {Springer},
  pages = {877-879},
  editor = {M.-Y. Kao},
  title = {Stable {M}arriage}
}

@article{Irv85,
  year = {1985},
  title = {An efficient algorithm for the ``stable roommates'' problem},
  pages = {577-595},
  author = {R.W. Irving},
  journal = {Journal of Algorithms},
  volume = {6},
    doi={10.1016/0196-6774(85)90033-1}
}

@article{IS07,
  year = {2007},
  title = {The Stable Fixtures Problem},
  pages = {2118-2129},
  author = {R.W. Irving and S. Scott},
  journal = {Discrete Applied Mathematics},
  volume = {155},
    doi = {j.dam.2007.05.015}
}

@conference{KIMS15,
  year = {2015},
  series = {Lecture Notes in Computer Science},
  title = {Profile-based optimal matchings in the {S}tudent--{P}roject {A}llocation problem},
  author = {A. Kwanashie and R.W. Irving and D.F. Manlove and C.T.S. Sng},
  publisher = {Springer},
    doi ={10.1007/978-3-319-19315-1_19},
  booktitle = {Proceedings of IWOCA 2014: the 25th International Workshop on Combinatorial Algorithms}
}

@inproceedings{Kir08,
  year = {2008},
  series = {Lecture Notes in Computer Science},
  title = {Better and simpler approximation algorithms for the stable marriage problem},
  pages = {623-634},
  author = {Z. Kir\'{a}ly},
  volume = {5193},
  publisher = {Springer},
  booktitle = {Proceedings of ESA '08: the 16th Annual European Symposium on Algorithms},
    doi={10.1007/978-3-540-87744-8_52}
}

@book{Knu76,
  year = {1976},
  title = {Mariages Stables et leurs relations avec d'autres probl\`emes combinatoires},
  author = {D.E. Knuth},
  note = {{E}nglish translation in \emph{Stable Marriage and its Relation to Other Combinatorial Problems}, volume 10 of CRM Proceedings and Lecture Notes, American Mathematical Society, 1997},
  publisher = {Les Presses de L'Universit\'{e} de Montr\'{e}al}
}

@incollection{Man08,
  booktitle = {Encyclopedia of Algorithms},
  year = {2008},
  author = {D.F. Manlove},
  publisher = {Springer},
  pages = {390-394},
  editor = {M.-Y. Kao},
  title = {Hospitals / {R}esidents problem}
}

@techreport{MI08,
  year = {2008},
  title = {Popular Matchings: Structure and Algorithms},
  institution = {University of Glasgow, Department of Computing Science},
  number = {TR-2008-292},
  author = {E. McDermid and R. Irving}
}

@book{networkflows,
author = {Ahuja, Ravindra K. and Magnanti, Thomas L. and Orlin, James B.},
title = {Network flows: theory, algorithms, and applications},
year = {1993},
isbn = {013617549X},
publisher = {Prentice-Hall, Inc.},
address = {USA}
}

@article{MW71,
  year = {1971},
  title = {The stable marriage problem},
  pages = {486-490},
  number = {7},
  author = {D.G. McVitie and L.B. Wilson},
    doi ={10.1145/362619.362631},
  journal = {Communications of the ACM},
  volume = {14}
}

@misc{matwa,
  url = {https://matwa.optimalmatching.com/},
  title = {MATWA},
    key={matwa}
}

@misc{matwamanual,
  url = {https://matwa.optimalmatching.com/matching/manual},
  title = {MATWA Guide},
    key={matwaguide}
}

@incollection{NRS08,
  booktitle = {The New Palgrave Dictionary of Economics},
  year = {2008},
  author = {M. Niederle and A.E. Roth and T. S\"onmez},
  publisher = {Palgrave-Macmillan},
  edition = {2},
  editor = {S. Derlauf and L. Blume},
  title = {Matching and market design}
}

@book{RS90,
  year = {1990},
  series = {Econometric Society Monographs},
  title = {Two-Sided Matching: a Study in Game-Theoretic Modeling and Analysis},
  author = {A.E. Roth and M.A.O. Sotomayor},
  volume = {18},
  publisher = {Cambridge University Press}
}

@article{RX94,
  year = {1994},
  title = {Jumping the gun: imperfections and institutions related to the timing of market transactions},
  pages = {992-1044},
  number = {4},
  author = {A.E. Roth and X. Xing},
  journal = {American Economic Review},
  volume = {84}
}

@article{SS74,
  year = {1974},
  title = {On cores and indivisibility},
  pages = {23-37},
  author = {L. Shapley and H. Scarf},
  journal = {Journal of Mathematical Economics},
  volume = {1},
doi = {10.1016/0304-4068(74)90033-0}
}

@article{Sub94,
  year = {1994},
  title = {A New Approach to Stable Matching Problems},
  pages = {671-700},
  number = {4},
  author = {A. Subramanian},
  journal = {SIAM Journal on Computing},
  volume = {23},
    doi={10.1137/S009753978916948}
}

@article{Tan91a,
  year = {1991},
  title = {Stable matchings and stable partitions},
  pages = {11-20},
  author = {J.J.M. Tan},
  journal = {International Journal of Computer Mathematics},
  volume = {39},
    doi={10.1080/00207169108803975}
}

@article{TH95,
  year = {1995},
  title = {A generalization of the stable matching problem},
  pages = {87-102},
  author = {J.J.M. Tan and Y.-C. Hsueh},
  journal = {Discrete Applied Mathematics},
  volume = {59},
    doi={10.1016/0166-218X(93)E0154-Q}
}

@book{matchUp,
  author    = {David Manlove},
  title     = {Algorithmics of Matching Under Preferences},
  series    = {Series on Theoretical Computer Science},
  volume    = {2},
  publisher = {World Scientific},
  year      = {2013},
  doi       = {10.1142/8591},
  isbn      = {978-981-4425-24-7},
  bibsource = {dblp computer science bibliography, https://dblp.org}
}

@article{ABRAHAM,
title = {Two algorithms for the Student-Project Allocation problem},
journal = {Journal of Discrete Algorithms},
volume = {5},
number = {1},
pages = {73-90},
year = {2007},
issn = {1570-8667},
doi = {https://doi.org/10.1016/j.jda.2006.03.006},
author = {David Abraham and Robert Irving and David Manlove}
}

@InProceedings{sofiatStrong,
author="Olaosebikan, Sofiat
and Manlove, David",
editor="Changat, Manoj
and Das, Sandip",
title="An Algorithm for Strong Stability in the Student-Project Allocation Problem with Ties",
booktitle="Algorithms and Discrete Applied Mathematics",
year="2020",
publisher="Springer International Publishing",
address="Cham",
pages="384--399",
isbn="978-3-030-39219-2",
doi={10.1007/978-3-030-39219-2_31}
}

@phdthesis{SofiatPhD,
    title    = {The Student-Project Allocation Problem: structure and algorithms},
    school   = {University of Glasgow},
    author   = {Olaosebikan, Sofiat},
    year     = {2020},
    url = {https://theses.gla.ac.uk/81514/}
}

@phdthesis{Augustine,
    title    = {Efficient algorithms for optimal matching problems under preferences},
    school   = {University of Glasgow},
    author   = {Kwanashie, Augustine},
    year     = {2015},
    url = {https://theses.gla.ac.uk/6706/}
}

@phdthesis{CooperPhD,
    title    = {Fair and large stable matchings in the stable marriage and student-project allocation problems},
    school   = {University of Glasgow},
    author   = {Cooper, Frances},
    url = {https://theses.gla.ac.uk/81622/},
    year     = {2020}
}

@Article{a6030471,
AUTHOR = {Király, Zoltán},
TITLE = {Linear Time Local Approximation Algorithm for Maximum Stable Marriage},
JOURNAL = {Algorithms},
VOLUME = {6},
YEAR = {2013},
NUMBER = {3},
PAGES = {471--484},
doi = {10.3390/a6030471},
ISSN = {1999-4893},
DOI = {10.3390/a6030471}
}

@mastersthesis{frederikbsc,
   author = {Glitzner, Frederik},
   title = {Student-Project Allocation in the Matching Algorithm Toolkit},
   year = {2023},
    school   = {University of Glasgow},
   type={BSc Computing Science Level 4 Project Dissertation},
    url = {https://glitznerf.github.io/files/level4project.pdf}
}

@InProceedings{profile,
author="Kwanashie, Augustine
and Irving, Robert 
and Manlove, David 
and Sng, Colin ",
editor="Jan, Kratochv{\'i}l
and Miller, Mirka
and Froncek, Dalibor",
title="Profile-Based Optimal Matchings in the Student/Project Allocation Problem",
booktitle="Combinatorial Algorithms",
year="2015",
publisher="Springer International Publishing",
address="Cham",
doi={10.1007/978-3-319-19315-1_19},
pages="213--225",
isbn="978-3-319-19315-1"
}

@techreport{greedymatchings,
    author = {Irving, Rob},
    title = {Greedy Matchings},
    institution = {University of Glasgow, Computing Science Department},
  note = {TR-2003-136},    
    url = {https://www.dcs.gla.ac.uk/~rwi/greedy.pdf},
    year = 2003 
}

@misc{TUM,
  title = {TUM Matching},
  author = {{Technical University of Munich}},
  howpublished = {\url{https://algorithms.discrete.ma.tum.de/matching/}},
  year={2016},
  note = {Accessed: 2024-08-01}
}

@misc{VisualMatchingAlgo,
  title = {VisuAlgo},
  author = {Steven Halim},
  howpublished = {\url{https://visualgo.net/en/matching}},
  year={2011},
  note = {Accessed: 2024-08-01}
}

@article{tanhsueh,
   author = {Jimmy J.M. Tan and Yuang-Cheh Hsueh},
   doi = {10.1016/0166-218X(93)E0154-Q},
   issn = {0166-218X},
   issue = {1},
   journal = {Discrete Applied Mathematics},
   month = {4},
   pages = {87-102},
   publisher = {North-Holland},
   title = {A generalization of the stable matching problem},
   volume = {59},
   year = {1995},
}

@misc{AlgMatch,
  title = {AlgMatch},
  author = {Liam Lau and Callum Ormond},
  howpublished = {\url{https://callumormond.github.io/individual-project/}},
  year={2023},
  note = {Accessed: 2024-08-01}
}

@misc{omalleytool,
  title = {{Optimal Matching Tool}},
  author = {Gregg O'Malley},
  howpublished = {\url{https://projects.optimalmatching.com/matchings}},
  year={2007},
  note = {Accessed: 2024-08-01}
}

@dataset{coopercode,
  author       = {Cooper, Frances},
  title        = {{Correctness tests for python package 
                   matchingproblems}},
  month        = oct,
  year         = 2020,
  publisher    = {Zenodo},
  version      = {1.0},
  doi          = {10.5281/zenodo.4065149}
}

@article{Hungarian,
  title={The Hungarian method for the assignment problem},
  author={Harold W. Kuhn},
  journal={Naval Research Logistics (NRL)},
  year={1955},
  volume={52}
}

@article{MatchU, title={MatchU: An Interactive Matching Platform}, volume={34}, DOI={10.1609/aaai.v34i09.7090}, abstractNote={&lt;p&gt;MatchU is a web-based platform that offers an interactive framework to find how to form mutually-beneficial relationships, decide how to distribute resources, or resolve conflicts through a suite of matching algorithms rooted in economics and artificial intelligence. In this paper, we discuss MatchU’s vision, solutions, and future directions.&lt;/p&gt;}, number={09}, journal={Proceedings of the AAAI Conference on Artificial Intelligence}, author={Ferris, James and Hosseini, Hadi}, year={2020}, month={4}, pages={13606-13607} }

@techreport{networkx,
  title={Exploring network structure, dynamics, and function using NetworkX},
  author={Hagberg, Aric and Swart, Pieter and S Chult, Daniel},
  year={2008},
  institution={Los Alamos National Lab (LANL), Los Alamos, NM (United States)}
}

@misc{cpsat,
  title = {CP-SAT Solver},
  author = {\mbox{Google Inc.}},
  howpublished = {\url{https://developers.google.com/optimization/cp/cp_solver}},
  note = {Accessed: 2024-08-01}
}

@article{metahr,
title = {A unified approach to finding good stable matchings in the hospitals/residents setting},
journal = {Theoretical Computer Science},
volume = {400},
number = {1},
pages = {84-99},
year = {2008},
issn = {0304-3975},
doi = {https://doi.org/10.1016/j.tcs.2008.02.014},
author = {Christine Cheng and Eric McDermid and Ichiro Suzuki}
}

@article{matchingpackage,
    doi = {10.21105/joss.02169},
    year = {2020},
    publisher = {The Open Journal},
    volume = {5},
    number = {48},
    pages = {2169},
    author = {Henry Wilde and Vincent Knight and Jonathan Gillard},
    title = {Matching: A Python library for solving matching games},
    journal = {Journal of Open Source Software}
}

@misc{lemon,
  title = {Library for Efficient Modeling and Optimization in Networks (LEMON)},
  author = {{Egerváry Research Group on Combinatorial Optimization, Eötvös Loránd University}},
  howpublished = {\url{https://lemon.cs.elte.hu/trac/lemon}},
  year={2003},
  note = {Accessed: 2024-08-01}
}

@misc{glpk,
  title = {GNU Linear Programming Kit (GLPK)},
  author = {\mbox{GNU project}},
  howpublished = {\url{https://www.gnu.org/software/glpk/glpk.html}},
  note = {Accessed: 2024-08-01}
}

@misc{RichardMorey,
  title = {Student project allocation},
  author = {Richard D Morey},
  howpublished = {\url{https://richarddmorey.github.io/studentProjectAllocation/}},
  year={2021},
  note = {Accessed: 2024-08-01}
}

@mastersthesis{PhilipYuile,
   author = {Yuile, Philip},
   title = {Adding to a Library of Matching Algorithms},
   year = {2011},
    school   = {University of Glasgow},
   type={BSc Computing Science Level 4 Project Dissertation}
}

@mastersthesis{Boris,
   author = {Lazarov, Boris},
   title = {A web app for visualising matching algorithms},
   year = {2018},
    school   = {University of Glasgow},
   type={BSc Computing Science Level 4 Project Dissertation}
}

@mastersthesis{AlesRemta,
   author = {Ales Remta},
   title = {A Java API for Matching Problems},
   year = {2010},
    school   = {University of Glasgow},
   type={MSc Computing Science Project Dissertation}
}

\clearpage

\appendix

\section{Appendix}
\subsection{Example API Usage}
\label{section:apiappendix}

\textit{MATWS} is a standalone REST API that provides three public endpoints to power \textit{MATWA}:
\begin{itemize}
 \item {\tt check-file} is called when the user chooses to upload their custom instances. It takes the problem class and instance strings as input and returns a status, instance strings, and a list of applicable algorithms. 
 \item {\tt check-params} is called when the user chooses to generate random instances. It takes the problem class and relevant parameters as input and returns a status, instance strings, and list of applicable algorithms just like {\tt check-file}.
 \item {\tt run-algorithms} is called when the user chooses a subset of the algorithms to apply to the instances. It takes the problem class, algorithm name, and instance strings as input, and returns a status, algorithm name, number of matchings found, list of those matchings, statistics, and graphs (if applicable).
\end{itemize}
\label{section:apiusage}

\subsubsection{Fetching Applicable Algorithms using {\tt check-file}}

\begin{apiRoute}{post}{https://matws.optimalmatching.com/check-file}{check custom instances and get applicable algorithms}
\begin{routeRequest}{application/json}
    \begin{routeRequestBody}
{
    "problemClass": "SR",
    "fileContents": "2\n2 \n1"
}
    \end{routeRequestBody}
\end{routeRequest}
\begin{routeResponse}{application/json}
    \begin{routeResponseItem}{201}{ok}
        \begin{routeResponseItemBody}
{
    "status": "success",
    "statusText": null,
    "availableAlgs": {
        "Minimum Regret Matching": "Find a minimum regret stable matching or report that none exists",
        (...)
        "Egalitarian Stable Matching": "Find an egalitarian stable matching or report that none exists"
    },
    "instances": null
}
        \end{routeResponseItemBody}
    \end{routeResponseItem}    
\end{routeResponse}
\end{apiRoute}

\clearpage
\subsubsection{Generating Instances using {\tt check-params}}

\begin{apiRoute}{post}{https://matws.optimalmatching.com/check-params}{generate random instances and get applicable algorithms}
\begin{routeRequest}{application/json}
    \begin{routeRequestBody}
{
    "problemClass": "SR",
    "parameters": {
        "numOfRoommates": 2,
        "probabilityOfTies": 0,
        "preferenceListDensity": 1,
        "numOfInstances": 1
    }
}
    \end{routeRequestBody}
\end{routeRequest}
\begin{routeResponse}{application/json}
    \begin{routeResponseItem}{201}{ok}
        \begin{routeResponseItemBody}
{
  "status": "success",
  "statusText": null,
  "availableAlgs": {
    "Minimum Regret Matching": "Find a minimum regret stable matching or report that none exists",
    (...)
    "Egalitarian Stable Matching": "Find an egalitarian stable matching or report that none exists"
  },
  "instances": [
    "2\n2 \n1"
  ]
}
        \end{routeResponseItemBody}
    \end{routeResponseItem}    
\end{routeResponse}
\end{apiRoute}

\clearpage
\subsubsection{Solving Instances using {\tt run-algorithms}}

\begin{apiRoute}{post}{https://matws.optimalmatching.com/run-algorithms}{run selected algorithms on given instances}
\begin{routeRequest}{application/json}
    \begin{routeRequestBody}
{
    "problemClass": "SR",
    "algorithms": "Default Stable (No Ties)",
    "fileContents": ["2\n2 \n1"]
}
    \end{routeRequestBody}
\end{routeRequest}
\begin{routeResponse}{application/json}
    \begin{routeResponseItem}{201}{ok}
        \begin{routeResponseItemBody}
[
    {
        "status": "success",
        "description": null,
        "algorithm": "default stable (no ties)",
        "numberOfMatchings": 1,
        "matchings": [
            {
                "matchingNumber": 0,
                "statsToDisplay": {
                    "Size": "1",
                    "Profile Amount": "(2)",
                    "Profile Position": "(1)",
                    "Cost": "2"
                },
                "statsToExpand": {
                    "Preference Lists": "1: <2>\n2: <1>",
                    "Matched Pairs": "(1, 2)\n"
                }
            }
        ],
        "numberOfIterations": 1,
        "stats": "",
        "expandableStats": {},
        "graphs": []
    }
]
        \end{routeResponseItemBody}
    \end{routeResponseItem}    
\end{routeResponse}
\end{apiRoute}

\clearpage
\subsection{Currently Available Algorithms}
\label{section:algorithms}
Below is a full list of the algorithms available in the system (as of August 2024) with references for further information. Note that some of these are only available for certain configurations of the problem model, so for a full colour-coded table to see, in more detail, which configuration admits which algorithms, please refer to the table linked in the \textcite{matwamanual}.

\paragraph{\sc Capacitated House Allocation}
\begin{itemize}
    \item \textbf{Naive:} Produces a matching using the random serial dictatorship mechanism. \cite{matchUp}
    \item \textbf{Minimum Cost:} Produces a minimum cost maximum matching. \cite{networkflows}
    \item \textbf{Rank-Maximal:} Produces a rank-maximal matching. \cite{IKMMP06}
    \item \textbf{Greedy:} Produces a greedy maximum matching. \cite{greedymatchings}
    \item \textbf{Generous:} Produces a generous maximum matching. \cite{greedymatchings}
    \item \textbf{Greedy-Generous:} Produces a greedy-generous maximum matching. \cite{greedymatchings}
    \item \textbf{Maximum Cardinality Pareto Optimal:} Produces a maximum cardinality Pareto optimal matching. \cite{ACMM04}
    \item \textbf{Popular:} Produces a popular matching or reports that none exists. \cite{AIKM05}
    \item \textbf{Switching Graph:} Visualises the switching graph for popular matchings. \cite{MI08}
\end{itemize}

\paragraph{\sc House Allocation}
\begin{itemize}
    \item \textit{(all algorithms for {\sc Capacitated House Allocation})}
    \item \textbf{Rank-Maximal Popular:} Produces a rank-maximal popular matching or reports that none exists. \cite{MI08}
    \item \textbf{Popular Uniform at Random:} Produces a popular matching uniform at random, or reports that none exists. \cite{MI08}
    \item \textbf{Generous Maximum Cardinality Popular:} Produces a maximum cardinality popular matching with a generous profile. \cite{MI08}
    \item \textbf{Minimum Cost Maximum Cardinality Popular:} Produces a maximum cardinality popular matching with minimum cost. \cite{MI08}
    \item \textbf{Popular Pairs:} Finds all admitted popular pairs. \cite{MI08}
    \item \textbf{Number of Popular Matchings:} Computes the number of popular matchings admitted by the instance. \cite{MI08}
    \item \textbf{All Popular Matchings:} Finds all admitted popular matchings. \cite{MI08}
\end{itemize}

\paragraph{\sc Hospitals / Residents}
\begin{itemize}
    \item \textbf{No-Ties Stable:} Produces the resident-optimal stable matching. \cite{GS62}
    \item \textbf{Super Stable:} Produces a super-stable matching, or reports that none exists. \cite{hrt}
    \item \textbf{Kiraly One-Sided Ties:} Produces an approximation for a maximum stable matching for instances with complete or incomplete lists and ties only on the hospital side. \cite{Kir08}
    \item \textbf{Kiraly Two-Sided Ties:} Produces an approximation for a maximum stable matching for instances with complete or incomplete lists and ties on both sides. \cite{a6030471}
\end{itemize}

\paragraph{\sc Stable Marriage}
\begin{itemize}
    \item \textit{(all algorithms for {\sc Hospitals / Residents})}
    \item \textbf{Maximum Popular:} Produces a maximum popular matching in the {\sc smi} context. \cite{Gus87}
    \item \textbf{Strongly Stable:} Produces a strongly stable matching in the {\sc smti} context, or reports that none exists. \cite{IMS02}
    \item \textbf{Egalitarian Stable:} Produces an egalitarian stable matching in the {\sc sm} context (requires complete preference lists). \cite{ILG87}
    \item \textbf{Minimum Regret Stable:} Produces a minimum regret stable matching in the {\sc sm} context (requires complete preference lists). \cite{Gus87}
    \item \textbf{Minimum $M$-Regret Stable:} Produces a stable matching in the {\sc sm} context with minimum regret over the first agent set ($M$) (requires complete preference lists). \cite{Gus87}
    \item \textbf{Minimum $W$-Regret Stable:} Produces a stable matching in the {\sc sm} context with minimum regret over the second agent set ($W$) (requires complete preference lists). \cite{Gus87}
    \item \textbf{All Stable Pairs:} Finds all stable pairs through enumeration in the {\sc sm} context (requires complete preference lists). \cite{Gus87}
    \item \textbf{All Stable Matchings:} Finds all stable matchings (two different algorithms for different settings, either Break-Marriage \cite{MW71} or Rotation-Elimination \cite{Gus87}).
\end{itemize}

\paragraph{\sc Stable Roommates}
\begin{itemize}
    \item \textbf{Tan-Hsueh:} Produces a reduced stable partition in a given instance with arbitrary tie-breaking in the presence of ties. \cite{TH95}
    \item \textbf{No-Ties Stable:} Produces a stable matching or reports that none exists. \cite{Irv85}
    \item \textbf{Maximum Stable:} Produces a maximum stable matching by deleting one agent from each odd cycle of a reduced stable partition. \cite{Tan91a}
    \item \textbf{Minimum Regret Stable:} Produces a minimum regret stable matching, or reports that none exists. \cite{GI89}
    \item \textbf{Egalitarian Stable:} Produces an egalitarian stable matching, or reports that none exists. \cite{gusfieldstructure}
    \item \textbf{All Stable Pairs:} Finds all (if any) admitted stable pairs through enumeration. \cite{gusfieldstructure}
    \item \textbf{All Stable Matchings:} Finds all (if any) admitted stable matchings through enumeration. \cite{gusfieldstructure}
\end{itemize}

\paragraph{\sc Student-Project Allocation}
\begin{itemize}
    \item \textbf{Cost-Optimal One-Sided:} Produces a minimum cost maximum matching considering only student preferences. \cite{KIMS15}
    \item \textbf{Greedy One-Sided:} Produces a greedy maximum matching considering only student preferences. \cite{KIMS15}
    \item \textbf{Generous One-Sided:} Produces a generous maximum matching considering only student preferences. \cite{KIMS15}
\end{itemize}

\paragraph{\sc Student-Project Allocation with Lecturer Preferences over Students}
\begin{itemize}
    \item \textit{(all algorithms for {\sc Student-Project Allocation})}
    \item \textbf{Student-Optimal Stable:} Produces the student-optimal stable matching. \cite{ABRAHAM}
    \item \textbf{Lecturer-Optimal Stable:} Produces the lecturer-optimal stable matching. \cite{ABRAHAM}
\end{itemize}

\end{document}